\documentclass[letterpaper,11pt]{article}
\usepackage{caption}
\pdfoutput=1 

\usepackage{extarrows}
\usepackage{amssymb}
\usepackage{amsmath}
\DeclareMathOperator{\Tr}{Tr}
\usepackage{amstext}
\usepackage{etoolbox}
\let\tinymatrix\smallmatrix

\patchcmd{\tinymatrix}{\scriptstyle}{\scriptscriptstyle}{}{}
\patchcmd{\tinymatrix}{\scriptstyle}{\scriptscriptstyle}{}{}
\patchcmd{\tinymatrix}{\vcenter}{\vtop}{}{}
\patchcmd{\tinymatrix}{\bgroup}{\bgroup\scriptsize}{}{}

\usepackage{graphicx,epsfig}
\usepackage{epsfig}
\usepackage{verbatim} 
\usepackage{color}
\usepackage{ulem}
\usepackage{enumitem}
\usepackage{subfigure}
\usepackage{bbm}
\usepackage{parskip}
\usepackage{dsfont}
\usepackage[numbers,sort&compress]{natbib}
\usepackage[body={6.5in, 9in}, right=1in, top=1in]{geometry}
\usepackage{mathtools}
\usepackage{comment}
\usepackage{caption}
\usepackage{bbold}
\usepackage{float}

\newcommand{\Comment}[1]{{}}
\definecolor{darkblue}{rgb}{0.15,0.35,0.55}
\definecolor{reddish}{rgb}{0.65, 0.2, 0.2}
\usepackage[linktocpage=true]{hyperref}

\newcommand{\be}{\begin{equation}}
\newcommand{\ee}{\end{equation}}
\newcommand{\bea}{\begin{eqnarray}}
\newcommand{\eea}{\end{eqnarray}}
\newcommand{\beas}{\begin{eqnarray*}}
\newcommand{\eeas}{\end{eqnarray*}}

\def\({\left(}
\def\){\right)}

\newcommand{\dT}{\mathcal{R}} 

\def\gsim{ \lower .75ex \hbox{$\sim$} \llap{\raise .27ex \hbox{$>$}} }
\def\lsim{ \lower .75ex \hbox{$\sim$} \llap{\raise .27ex \hbox{$<$}} }

\usepackage{xcolor,colortbl}

\begin{document}
\def\thefootnote{\fnsymbol{footnote}}

\begin{center}
\LARGE{\textbf{Search Optimization, Funnel Topography, and Dynamical Criticality on the String Landscape}} \\[0.5cm]
 
\large{Justin Khoury and Onkar Parrikar}
\\[0.5cm]

\small{
\textit{Center for Particle Cosmology, Department of Physics and Astronomy, University of Pennsylvania,\\ 209 South 33rd St, Philadelphia, PA 19104}}

\vspace{.2cm}

\end{center}

\vspace{.6cm}

\hrule \vspace{0.2cm}
\centerline{\small{\bf Abstract}}
{\small\noindent A striking feature of our universe is its near criticality. The cosmological constant and weak hierarchy problems,
as well as the metastability of the electroweak vacuum, can all be understood as problems of criticality. This suggests a statistical
physics approach, based on the landscape of string theory. In this paper we present a dynamical selection mechanism for hospitable
vacua based on search optimization. Instead of focusing on late-time, stationary probability distributions for the different vacua, we
are interested in the approach to equilibrium. This is particularly relevant if cosmological evolution on the multiverse has occurred for a finite time much shorter than the exponentially-long mixing time for the landscape. We argue this imposes a strong selection pressure among hospitable vacua, favoring those that lie in regions where the search algorithm
is efficient. Specifically, we show that the mean first passage time is minimized for hospitable vacua lying at the bottom of funnel-like regions, akin to the smooth folding funnels of naturally-occurring
proteins and the convex loss functions of well-trained deep neural networks. The optimality criterion is time-reparametrization invariant and defined by two competing requirements: search efficiency, which requires minimizing the mean first passage time, and sweeping exploration, which requires that random walks are recurrent. Optimal landscape regions reach a compromise by lying at the critical boundary between recurrence and transience, thereby achieving dynamical criticality. Remarkably, this implies that the optimal lifetime of vacua coincides with the de Sitter Page time, $\tau_{\rm crit} \sim M_{\rm Pl}^2/H^3$. Our mechanism makes concrete phenomenological predictions: 1.~The expected lifetime of our universe is $10^{130}~{\rm years}$, which is $\gsim\; 2\sigma$ from the Standard Model metastability estimate; 2.~The supersymmetry breaking scale should be high, $\gsim\; 10^{10}~{\rm GeV}$. The present framework suggests a correspondence between the near-criticality of our universe and non-equilibrium critical phenomena on the landscape.
\vspace{0.3cm}
\noindent
\hrule
\def\thefootnote{\arabic{footnote}}
\setcounter{footnote}{0}

\section{Introduction}

Naturalness and symmetry have been the guiding principles of particle physics for the last fifty years. 
They were central in the development of the Standard Model (SM) and have steered Beyond-the-SM (BSM) model building. It appears increasingly doubtful, however, that these cherished principles 
can explain the fine-tunings currently facing theoretical physics. The lack of evidence for new physics at the Large Hadron Collider (LHC) suggests that Nature may have chosen the most
parsimonious UV completion for the SM --- a weakly coupled scalar field,
with nothing else to stabilize the weak scale. The failure of direct detection experiments to discover Weakly Interacting Massive Particle (WIMP) dark matter, a natural by-product of low-scale supersymmetry (SUSY), only strengthens the case against new weak-scale physics. Last but not least is the cosmological constant, which remains steadfastly
immune to any compelling natural explanation~\cite{Weinberg:1988cp}.

Perhaps we have been unlucky. New particles might lie just beyond the reach of the LHC, and WIMPs may be
 buried beneath the neutrino floor. But the current state of affairs has led some to ponder whether particle physics has entered a 
{\it post-naturalness era}~\cite{Giudice:2017pzm}.

A disturbing consequence of a grand ``desert" above the weak scale is the metastability of the electroweak
vacuum~\cite{Frampton:1976kf,Sher:1988mj,Casas:1994qy,Espinosa:1995se,Isidori:2001bm,Espinosa:2007qp,Ellis:2009tp,Degrassi:2012ry,Buttazzo:2013uya,Lalak:2014qua,Andreassen:2014gha,Branchina:2014rva,Bednyakov:2015sca,Iacobellis:2016eof,Andreassen:2017rzq}. Within the SM the Higgs quartic coupling $\lambda(\mu)$
becomes negative at high energy, implying that our vacuum is metastable. The estimated lifetime of our vacuum is~\cite{Andreassen:2017rzq}\footnote{To be clear,~\eqref{tdecay obs intro} is the characteristic time to form a bubble a true vacuum within our observable universe.}
\be
\tau = 10^{526^{+409}_{-202}}~{\rm years}\,.
\label{tdecay obs intro}
\ee
While much longer than the age of the universe, it is an uncomfortably close call. The quoted lifetime hinges on a delicate cancellation between an exponentially small number --- the bounce factor, $\sim \exp\left(-8\pi^2/3|\lambda(\mu_\star)|\right)$, which is exquisitely sensitive to $\lambda$ at the scale $\mu_\star \simeq 3\times 10^{17}~{\rm GeV}$ where it achieves a minimum; and an exponentially large number --- the space-time volume of our universe.\footnote{The calculation of the tunneling rate per unit volume is done in flat space-time. Gravitational corrections are consistently small~\cite{Isidori:2007vm,Salvio:2016mvj}.} What makes the Higgs metastability particularly intriguing is that it relates the cosmological constant, which sets the characteristic 4-volume, and the weak scale, which sets the Higgs and top quark masses.  

We believe this numerical conspiracy is no accident. It begs the question --- {\it why is our universe so precariously close to the edge of instability?} Anthropic reasoning, which has been successfully applied to the vacuum energy~\cite{Weinberg:1987dv} and the Higgs expectation value~\cite{Agrawal:1997gf}, does not seemingly offer any guidance, for it is difficult to conceive {\it a priori} how vacuum metastability is a pre-requisite for the existence of observers. (See~\cite{Espinosa:2007qp,Kobakhidze:2013tn,Enqvist:2013kaa,Fairbairn:2014zia,Enqvist:2014bua,Herranen:2014cua,Kamada:2014ufa,Salvio:2015cja,Shkerin:2015exa,Kearney:2015vba,Espinosa:2015qea,Figueroa:2015rqa,Herranen:2015ima,Ballesteros:2016xej,Joti:2017fwe,Espinosa:2017sgp,Han:2018yrk,Salvio:2018rv} for cosmological implications of the Higgs metastability.)
 
The metastability of our vacuum can be interpreted as a problem of near-criticality~\cite{Buttazzo:2013uya}.\footnote{For instance, the decay rate per space-time volume is relatively close to the critical range for percolation~\cite{Guth:1982pn}.} Remarkably, other fine-tuning problems can also be viewed through the prism of criticality. Consider the weak hierarchy problem. As remarked in~\cite{Giudice:2006sn}, if the Higgs mass spans the natural range $-M_{\rm Pl}^4 \;\lsim\; m_h^2 \;\lsim\; M_{\rm Pl}^4$, then the puzzle of the measured $|m_h^2| \ll M_{\rm Pl}^2$ is that it is so close to the critical point between an unbroken phase with $m_h^2 \sim M^2$ and $\langle h \rangle \simeq 0$, and a badly broken phase with $m_h^2 \sim -M^2$ and $\langle h \rangle \sim M_{\rm Pl}$.\footnote{One should emphasize that even with $m_h^2 > 0$, the electroweak phase is spontaneously broken at the QCD scale by the Higgs coupling to the quark condensate. See~\cite{ArkaniHamed:2005yv,Samuel:1999am} for discussions of this other phase of the SM.}

Less well-defined but equally intuitive is the cosmological constant, whose minute value implies a nearly-flat universe. In some sense, Minkowski space represents a quantum critical point between de Sitter (dS) and anti-de Sitter (AdS) space-times, which have different asymptotics and dynamical stability properties~\cite{Friedrich:1987,Bizon:2011gg}. Last but not least is the near scale-invariance of primordial density perturbations, whose spectral index is suggestive of a critical exponent. The mechanism traditionally invoked to generate these perturbations, namely slow-roll inflation, itself represents a near-critical phenomenon, with the inflaton interpolating between a nearly de Sitter phase (approximate conformal fixed point) and standard decelerating expansion. 
 
 \subsection{Landscape approach}

The near criticality of our universe strongly suggests a statistical physics approach, based on the idea that the parameters of the SM vary across a vast landscape of
vacua, with the observed values determined environmentally. This is motivated by the exponentially large number of metastable states in string theory~\cite{Bousso:2000xa,Kachru:2003aw,Susskind:2003kw,Douglas:2003um}, together with the mechanism of eternal inflation~\cite{Vilenkin:1983xq,Linde:1986fc,Linde:1986fd} to dynamically populate these vacua.\footnote{Recently,  the existence of metastable dS vacua in string theory has been questioned~\cite{Obied:2018sgi,Agrawal:2018own}, which has sparked a heated debate ({\it e.g.}, see~\cite{Heckman:2018mxl} and references therein). In this paper we will assume that the fundamental theory admits a rich landscape of vacua, including metastable dS vacua.}

A significant challenge to this approach is the {\it measure problem}~\cite{Freivogel:2011eg}. Since bubbles of all type are generated infinitely-many times in an eternally inflating universe,
a regularization prescription ({\it i.e.}, a measure) is required to count bubbles and define late-time probabilities. There are two broad classes of measures. The first class comprises {\it global} measures: given a global foliation specified by some global time coordinate $t$, one counts bubbles of different types on a late-time hypersurface $t = t_{\rm c}$, and then let $t_{\rm c}\rightarrow \infty$~\cite{Linde:1993nz,Linde:1993xx,GarciaBellido:1993wn,Vilenkin:1994ua,Garriga:1997ef,Garriga:2005av,Vanchurin:2006qp,DeSimone:2008bq}. The resulting measure is independent of initial conditions, reflecting the attractor behavior of eternal inflation. A major drawback, however, is that it depends sensitively on the choice of global time variable~\cite{Linde:1993nz,Linde:1993xx,GarciaBellido:1993wn}.

The second class includes {\it local} measures, which focus on a space-time region around time-like geodesics. Examples include the past light-cone of world-lines (causal patch or causal diamond measure~\cite{Bousso:2006ev,Bousso:2009dm}), a region bounded by the apparent horizon~\cite{Bousso:2010zi}, or a space-like region around a world-line (``fat geodesic"~\cite{Bousso:2008hz}). In this paper we will primarily follow the ``watcher" measure, defined by an ensemble of time-like geodesics or ``watchers"~\cite{Garriga:2005av,Vanchurin:2006qp,Garriga:2012bc,Nomura:2011dt}. For instance, the ensemble can be generated as follows~\cite{Nomura:2011dt}: starting with an observer in some initial dS vacuum, we follow all future ``decohered" classical histories, thus generating an ensemble of world-lines weighted unambiguously by their quantum-mechanical branching ratios. 

A feature common to all local measures is that a typical geodesic will eventually enter an AdS or Minkowski vacuum. An AdS region collapses within a
Hubble time to a big crunch singularity, hence AdS vacua are usually assumed terminal. Minkowski vacua are also generally assumed to be terminal. Therefore, all
but a measure zero of watchers will sample a finite number of bubbles. This regulates the infinities of eternal inflation~\cite{Bousso:2006ev}, but introduces a dependence on initial conditions. In this work AdS and Minkowski vacua will be treated as terminal, and we will argue in Sec.~\ref{init cond} that under certain assumptions
our results are largely independent of initial conditions.

Underlying these approaches to the landscape is the assumption that the multiverse has existed for a sufficiently long time, such that vacuum statistics have nearly settled to a stationary/equilibrium distribution. To be precise, denote by $f_i(t)$ the probability that a watcher occupies vacuum $i$ at time $t$. (Equivalently, $f_i(t)$ is the fraction of comoving volume occupied by vacuum $i$.) Asymptotically, this tends to a stationary distribution $\vec{f}^{\,(0)}$, which lies entirely in the subspace of terminal vacua, up to an exponentially-decaying piece $\delta \vec{f} = \vec{s} e^{q t}$, which is dominated by the longest-lived dS vacuum, called the {\it dominant vacuum}. Since the latter is unlikely to be anthropically hospitable, the frequency of hospitable
vacua is determined by the tunneling rate from the dominant vacuum into hospitable ones. The hope is that, according to this measure, a universe like ours is statistically likely among all possible anthropically hospitable vacua. In other words, our universe ought to be generic or typical according to the measure. This is the {\it principle of mediocrity}. 

The assumption that the multiverse has nearly settled into a stationary state is non-trivial. As with many complex systems~\cite{evolution_complex}, in particular spin glasses~\cite{glassylandscape}, the landscape features many long-lived metastable vacua, resulting in frustration and aging dynamics. Consequently the mixing time for the landscape globally is exponentially long~\cite{Denef:2011ee}. In the language of computational complexity, the number of vacua $N$ scales exponentially with the effective dimensionality $D$ of the landscape, that is, $N \sim {\rm e}^D$, and the problem of finding a vacuum within a specified hospitable range of vacuum energy has been argued to be~\textsf{NP}-hard~\cite{Denef:2006ad}. See~\cite{Halverson:2018cio} for further implications. 

\begin{figure}[h]
\centering   
\subfigure[Fiducial region]{\label{region intro}\includegraphics[height=2.0in]{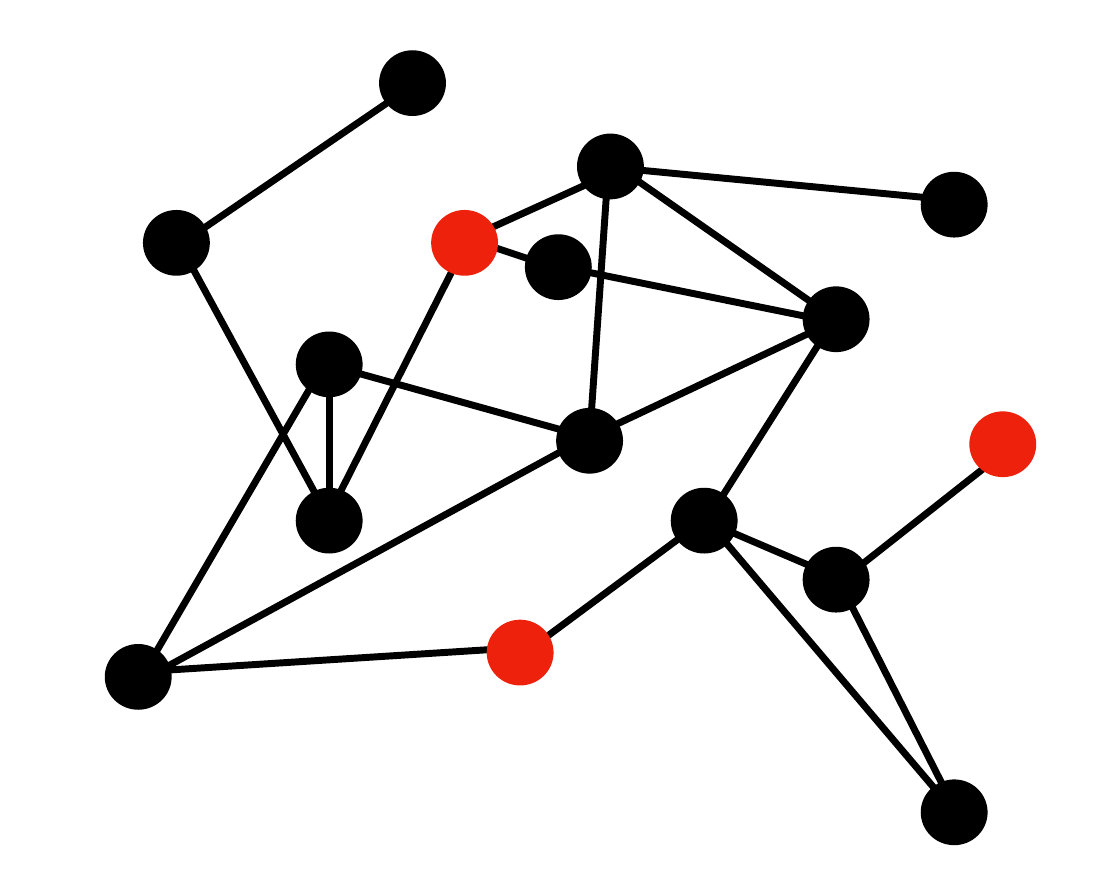}}
\hspace{0.75in}
\subfigure[Replicas with small variations]{\label{region copies intro}\includegraphics[height=2.0in]{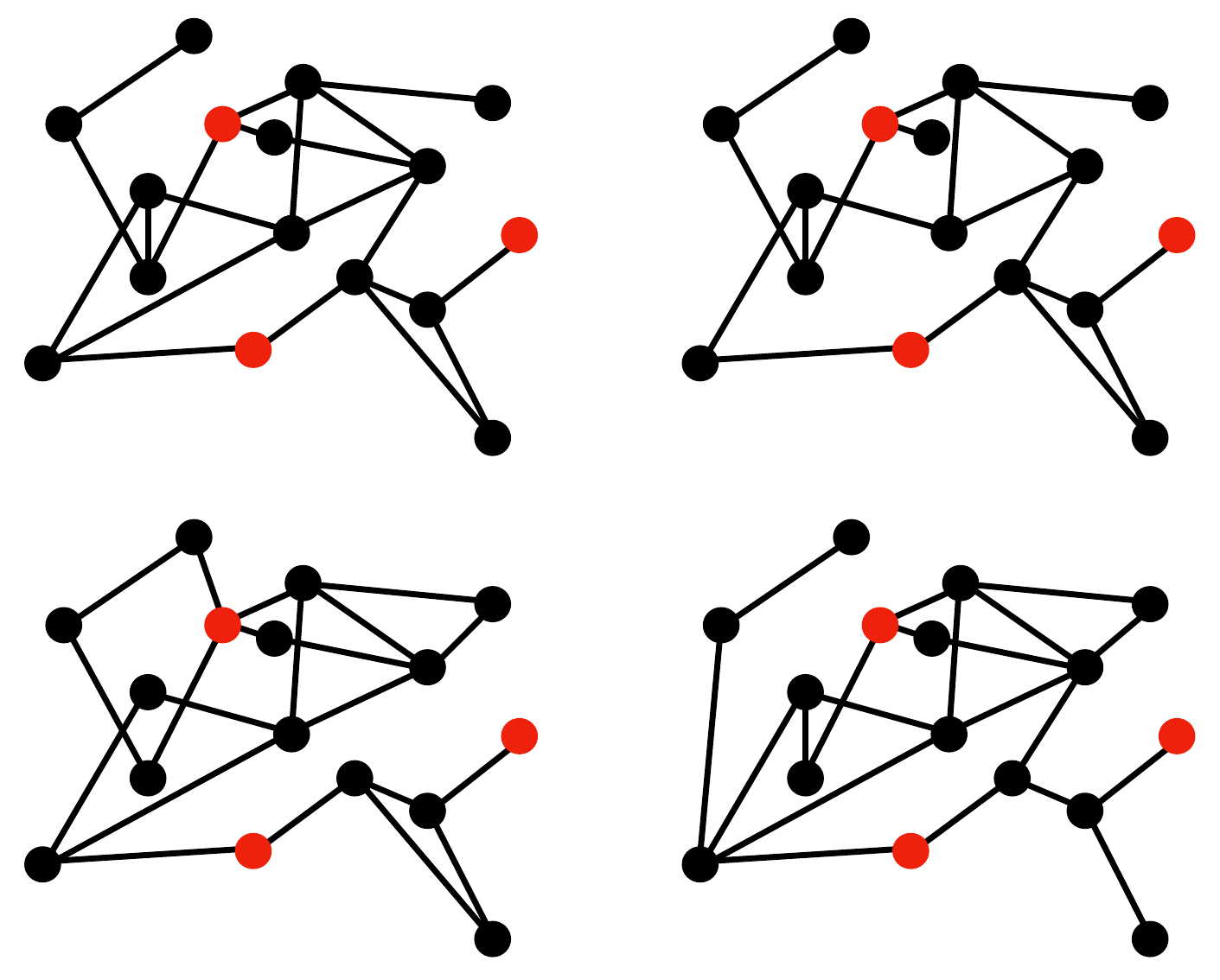}}
\caption{{\it a)} A fiducial region of the landscape, modeled as a weighted, undirected network. Nodes are metastable dS vacua, while links are the relevant transition rates. The red nodes are hospitable vacua. Tunneling rates into AdS/Minkowski vacua (not shown) introduce a rate of death. {\it b)} In the vastness of the landscape, we imagine there are many replicas of the fiducial region, each with slight variations. This is illustrated here with small changes in the network topology.  More generally, we also expect variations in other physical parameters.}
\label{network illustration}
\end{figure}

\subsection{Search optimization and funnel topography}

Instead of focusing on equilibrium distributions, our primary interest in this paper lies in the {\it approach to equilibrium}.
This is particularly relevant if, as observed by Denef {\it et al.}~\cite{Denef:2017cxt}, the multiverse has only existed for a finite time
much shorter than the exponentially-long mixing time for the landscape. Specifically we assume that all watchers in our ensemble hit
terminal vacua on a time scale much shorter than the landscape mixing time.\footnote{This effectively imposes an ``end-of-time" geometric
cutoff~\cite{Bousso:2010yn} at a time much shorter than the mixing time.} The finite-time perspective changes the relevant question for the likelihood of our vacuum. Instead of asking, What type of hospitable vacua occurs most frequently at steady-state according to the equilibrium probability distribution?, we are interested in the question --- What hospitable
vacua have the right properties to be easily accessed in the evolution~\cite{Denef:2017cxt}?\footnote{Like~\cite{Denef:2017cxt},
we remain agnostic as to whether hospitable vacua end up harboring observers or not. Observers will not play a central role in our discussion.} 

This translates to a search optimization problem: {\it vacua that are easily accessed reside in optimal regions where
the standard search algorithm is particularly efficient.}\footnote{The link between optimization algorithms and the cosmological constant was also explored in~\cite{Bao:2017thx}.} 
A goal of this paper is to make precise the conditions for optimality and derive phenomenological implications for hospitable vacua lying in optimal regions of the landscape. As we will see, the optimality criteria will be independent of the choice of time variable, as well as whether vacua are weighted by comoving or physical
volume. Furthermore, we will show that optimal regions are characterized by critical dynamics, thereby suggesting a connection
between the near-criticality of our universe and non-equilibrium critical phenomena on the landscape.

To be clear, the approach does not claim to solve the measure problem. Ultimately the likelihood for our vacuum to reside in an optimal region
{\it vs} a sub-optimal one hinges on the choice of measure. We will try to carefully delineate which aspects of the calculation are sensitive to the measure,
and which are not. 

Consider a finite region of the landscape comprised of $N\gg 1$ dS vacua, large enough to include $N_{\rm target} \ll N$ hospitable vacua.\footnote{Although it is helpful to cast the discussion initially in terms `hospitable' and `inhospitable' vacua, our natural selection mechanism will lead to phenomenological predictions that rely as little as possible on anthropic reasoning.} We only explicitly keep track of dS vacua --- the only role of AdS/Minkowski terminals is to introduce a ``death" rate for watchers. In other words, the occupational probabilities $f_i(t)$ are conditional on the watcher not having transitioned yet to a terminal vacuum and died. Treating the region as a closed system for the moment, and ignoring bubble collisions, the $f_i(t)$'s satisfy a linear Markov process~\cite{Garriga:1997ef,Garriga:2005av,Vanchurin:2006qp}, with transition rates governed by Coleman-De Luccia (CDL) instantons~\cite{Coleman:1977py,Callan:1977pt,Coleman:1980aw}. 

\begin{figure}[h]
\centering   
\subfigure[Generic, frustrated region]{\label{frustrated region}\includegraphics[height=2.0in]{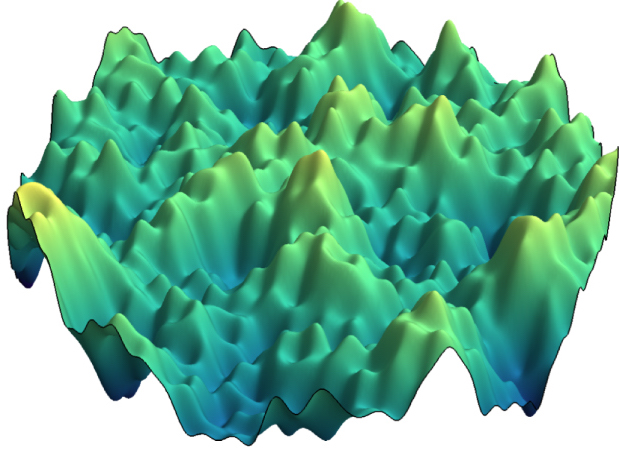}}
\hspace{0.75in}
\subfigure[Optimal, funnel-like region]{\label{optimal region}\includegraphics[height=2.25in]{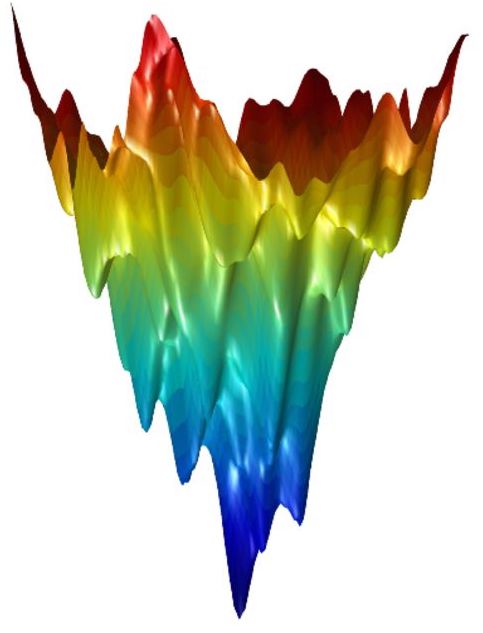}}
\caption{{\it a)} A generic region in the landscape typically has multiple vacua whose only allowed transitions are upward jumps. Because upward jumps have exponentially suppressed transition rates, the dynamics are frustrated, resulting in exponentially long search times. (Credit: Chiara Cammarota.) {\it b)} An optimal region is characterized by all vacua having at least one allowed downward transition, except for the lowest-lying vacuum. Its landscape topography is that of a funnel, akin to the free energy landscape of proteins and the loss function of well-trained deep neural networks. (Credit: Reproduced from~\cite{funnelfig}.)}
\label{generic vs optimal}
\end{figure}

Thus the region can be modeled as a weighted and undirected network of $N$ nodes, with the set of links specifying the topology of the network (Fig.~\ref{region intro})~\cite{Carifio:2017nyb}. 
As his/her space-time neighborhood undergoes a sequence of transitions, the watcher is performing a random walk on this network (Sec.~\ref{RW complex nets}). 
The subject of random walks on graphs (or networks) is a venerable branch of mathematics, but in the natural sciences it has enjoyed an explosion of activity in recent years~\cite{complex network PRL,complex network book,complex network review}. This is in large part due to their relevance to a myriad applications, from biological networks, to the internet, to epidemic propagation. 

In the vastness of the landscape, we imagine that our ensemble of watchers will sample an ensemble of landscape regions, comprised of many replicas of the fiducial region, each with slight variations (Fig.~\ref{region copies intro}). For concreteness, we assume all regions have on average $N={\rm e}^D$ vacua\footnote{Recall that $D$ is the dimensionality of the landscape, controlled by the number of underlying moduli.} and the same underlying statistical distribution of potential energies, but otherwise differ slightly from each other in their network topology and transition rates. We will argue that within a {\it generic} representative region of the ensemble, the average time to access hospitable vacua likely scales exponentially with $D$ --- consistent with the~\textsf{NP}-hard complexity class of the general problem. However, assuming the ensemble is large enough, there will be exceptional regions where the search for hospitable vacua is optimal, characterized by polynomial search time. 

Quantitatively, our figure of merit is a dimensionless {\it mean first-passage time} (MFPT)~\cite{MFPT book}, a widely-used measure of random walk efficiency~\cite{MFPT1,MFPT2,MFPT3}. The average MFPT, ${\cal T}_{\rm MFPT}$, also known as Kemeny's constant~\cite{kemeny}, quantifies the average time taken by a random walker to reach a target vacuum, randomly-selected according to the stationary distribution. As reviewed in Sec.~\ref{MFPT section} the average MFPT admits a simple expression as a spectral sum over the eigenvalues of the transition matrix. The result is, however, most intuitively stated in the ``downward" approximation~\cite{SchwartzPerlov:2006hi,Olum:2007yk}, which neglects the exponentially-small ``upward" CDL transitions~\cite{Lee:1987qc}:
\be
{\cal T}_{\rm MFPT}  \simeq \sum_{i\neq {\rm lowest}}^N \frac{1}{\kappa_i\Delta t}\,.
\label{MFPT intro}
\ee
Here, $\Delta t$ is a unit global time step, $\kappa_i$ is the total rate out of vacuum $i$, and the sum excludes the lowest-lying vacuum which is necessarily stable in the downward approximation. Remarkably,~\eqref{MFPT intro} is identical to the mean transport time in quenched disordered media~\cite{disordered media}. The result is intuitively clear --- the MFPT is the sum over the characteristic number of waiting time steps, $\left(\kappa_i\Delta t\right)^{-1}$, for all metastable vacua in the region. Importantly, since $\kappa_i$ is the decay rate per unit time $t$, the dimensionless MFPT is manifestly time-reparametrization invariant, {\it i.e.}, independent of the choice of~$t$. Furthermore, a well-known and remarkable property of Kemeny's constant is that {\it it is independent of the initial node for the random walk.} 

Now, a generic region in the ensemble (sketched in Fig.~\ref{frustrated region}) will typically include one or more metastable vacua whose only possible transitions involve up-tunneling. In the strict downward approximation, these vacua are absolutely stable, but more precisely at sub-leading order their decay rate is exponentially suppressed. The average MFPT in such generic regions is exponentially long. A watcher is overwhelmingly likely to tunnel to an AdS (or Minkowski) terminal well before accessing a hospitable vacuum. If the ensemble is large enough, however, there ought to be rare but optimal regions, characterized by all vacua having at least one allowed downward transition to another dS vacuum (except of course for the lowest-lying vacuum). As sketched in Fig.~\ref{optimal region}, optimal regions have the topography of a {\it broad valley or funnel}. Furthermore, to minimize~\eqref{MFPT intro} downward transitions should be as fast as possible (though importantly, as we will see later, transition rates are bounded from above by other considerations). This results in power-law search time. 

This is strikingly reminiscent of the protein folding problem~\cite{proteins1,proteins2}. Like the string landscape, the number of protein conformations (metastable vacua) scales exponentially with the number of amino acids (degrees of freedom). Correspondingly, the problem of finding the minimum energy state (so-called native state) is \textsf{NP}-complete~\cite{proteinNP}. Physically, generic sequences of amino acids get trapped in long-lived metastable conformations, resulting in exponentially-long relaxation time. Yet naturally-occurring proteins fold efficiently to their native state. This is Levinthal's paradox~\cite{Levinthal}. Nature's solution is of course evolution. The energy landscape of real proteins is characterized by high-energy unfolded states connected to the lowest-energy native state by a relatively smooth funnel.  This is known as the {\it principle of minimal frustration}~\cite{proteins1}.
 
Deep learning offers a similar narrative. Finding the global minimum of the loss function of deep neural networks has long been known to be~\textsf{NP}-complete~\cite{DNN NP}. Recent improvements in visualization techniques have revealed a tantalizing connection between network architecture and loss function topography~\cite{DNNvisual}. Deep neural networks with poor training parameters (or overtrained networks) have a highly non-convex landscape, featuring many local minima, whereas well-trained networks that easily generalize correspond to smooth, funnel-like landscapes. Furthermore, it has been shown recently that the connectivity matrix of well-trained, state-of-the-art deep neural networks have power-law spectral density~\cite{DNNpowerlaw}, which is well-described by heavy-tailed random matrix theory~\cite{RMT}. This indicates that well-trained networks operate at criticality. 
The connection between search optimization and criticality will be a recurring theme in our discussion.

\subsection{Recurrence {\it vs} transience}

We have so far approximated regions in the ensemble as closed systems, ignoring the exchange of probability with their surroundings. 
More realistically, regions are of course open systems, which allows the possibility that a random walker will escape a region {\it before} landing on a target vacuum. Optimal regions should therefore reach a comprise between efficient sampling, which tends to minimize the MFPT but increases the likelihood of escape, and oversampling, which minimizes the chance of
escape but results in sub-optimal~MFPT.

Treating regions as open systems would require modeling their environment, which in turn may  
introduce unwanted model-dependence in our analysis. Alternatively we propose to study a
proxy requirement that relies solely on the intrinsic dynamics within a given region. Specifically,
we consider the hypothetical limit of infinite volume, $N\rightarrow \infty$, and demand that 
random walks are {\it recurrent} in this limit. In recurrent walks, random walkers are certain to return to their starting point,
and will do so infinitely-many times in the future. Equivalently, in recurrent walks every site will be visited with probability one.
In contrast, for {\it transient} walks there is a finite probability that random walkers never return to their original site. 

To be clear, at the end of the day our ensemble still consists of finite-$N$ regions. But the behavior of
random walks in the $N\rightarrow \infty$ limit informs us on the potential impact of the environment of regions. 
Although not technically equivalent to modeling regions as open systems, we view the requirement of
recurrence as a reliable proxy for efficient sampling of regions. Recurrent walks efficiently
explore any region around their starting point, whereas transient walks tend to escape to infinity.\footnote{A classic result in the theory of random walks is P\'olya's theorem, which states that a simple random walk on a regular $d$-dimensional lattice is recurrent for $d \leq 2$, and transient for $d > 2$~\cite{polya}. As mathematician Shizuo Kakutani famously quipped, ``A drunken man will find his way home, but a drunken bird may get lost forever."} 

In Sec.~\ref{rec vs tran} we review the criterion for recurrence and show that it can be phrased in terms of a parameter~$\dT$, which is simply related to the average MFPT:
\be
\dT \equiv \frac{{\cal T}_{\rm MFPT}}{N}\,.
\ee
This can be interpreted as a mean residency time. Random walks are recurrent if~$\dT$ diverges as $N\rightarrow \infty$.
In the downward approximation, this criterion reduces to
\be
\dT \simeq \left\langle \frac{1}{\kappa_i\Delta t}\right\rangle \xrightarrow[N\rightarrow\infty]{} \;\infty \qquad ({\rm recurrence}) \,,
\label{Tau kappa}
\ee
where $\langle \ldots \rangle$ denotes the average over all vacua in the region. It should be intuitively clear that~\eqref{Tau kappa} is sensitive to the decay rate of the lowest-lying, and generically most stable, vacua. Random walks will be recurrent if the average decay rate diverges sufficiently fast as $V\rightarrow 0$.   

\subsection{Dynamical criticality and natural selection}

We now come to a key result of this paper. Search optimization is defined by two competing requirements: search efficiency, which requires fast transition rates 
to minimize the average MFPT, and sweeping exploration, which requires sufficiently long-lived vacua to achieve recurrence. Optimal regions reach a compromise
by having the shortest MFPT compatible with recurrence, {\it i.e.},  $\dT$ should diverge as $N\rightarrow\infty$ but at the slowest possible rate. In other words,
optimal regions lie at the critical boundary between recurrence and transience.

We will show in Sec.~\ref{edge of chaos} that optimality is achieved in regions characterized by vacua with average proper lifetime scaling as 
\be
\tau_{\rm crit} (V) \sim \frac{M_{\rm Pl}^5}{V^{3/2}}\sim \frac{M_{\rm Pl}^2}{H^3}~~~{\rm as}~V \rightarrow 0\,.
\label{tau crit intro}
\ee
Remarkably, this optimal lifetime coincides with the dS Page time~\cite{Page:1993wv}. The dS Page time has appeared in other inflationary studies before~\cite{Danielsson:2002td,Danielsson:2003wb,Ferreira:2016hee,Ferreira:2017ogo}. In slow-roll inflation, for instance, it marks the phase transition to eternal inflation~\cite{Creminelli:2008es,ArkaniHamed:2007ky}. The connection between the dS Page time and dynamically critical random walks on the landscape is surprising, and we do not yet have a compelling physical explanation for its occurrence.  

The optimal lifetime~\eqref{tau crit intro} implies that $\dT$ diverges as $\log N$, which signals dynamical criticality. A similar non-equilibrium phase transition occurs in quenched disordered media, when the probability distribution for waiting times reaches a critical power-law~\cite{disordered media}. The recurrence/transience boundary can be interpreted as critical from two points of view. Firstly, from a computational perspective, it represents a transition in the scaling of the recurrence measure $\dT$ as a function of the effective dimensionality $D = \ln N$. Recurrence typically results in a power-law divergence in $N \sim {\rm e}^D$, and hence exponential in $D$. The critical case instead diverges as $\log N \sim D$, and therefore polynomially in~$D$. More precisely, the large $N$ behavior of $\dT$ delineates regions in the landscape into three basic types:
\be
\dT  \sim \left\{\begin{array}{ccl}
\text{Exp}(D)&\cdots & \text{(recurrent but high complexity)} \\
 \text{Poly}(D)&\cdots & \text{(recurrent and low complexity)} \\
 \text{Constant}& \cdots &  \text{(low complexity but transient)}\,,
\end{array}\right.
\ee
where the second line corresponds to a critical boundary between the other two phases. The phase transition of interest is akin to {\it computational phase transitions}, which have the subject of much activity in recent years~\cite{compPT1,compPT2}. Secondly, from a dynamical perspective, it delineates regions of stability (recurrence) and instability (transience). Indeed, a pair of random walks starting from the same site will meet infinitely-many
times in the recurrent case, but are certain to eventually never meet again in the transient case. 

Thus the joint demands of sweeping exploration, defined by recurrence, and minimal oversampling, defined by minimal MFPT, select regions of the landscape that are tuned at criticality.
As such our mechanism realizes a form of {\it self-organized criticality}~\cite{SOC1,SOC2,SOC3}. It is consistent with the {\it dynamical criticality hypothesis}~\cite{dynamical crit review},
which states that complex systems maximize their computational capabilities at the phase transition between stable and unstable dynamical behavior. This idea goes back to random
boolean networks~\cite{Kauffman} and cellular automata~\cite{game of life,wolfram}. Specifically, cellular automata with certain ``edge of chaos"
dynamical rules (Class IV per Wolfram's classification) are capable of universal computation, exhibiting long-lived and complex transient structures~\cite{edge of chaos 1,edge of chaos 2,edge of chaos 3,edge of chaos 4,edge of chaos 5}. Recurrent neural networks~\cite{RNNchaos1,RNNchaos2}, specifically echo state networks~\cite{echo state}, achieve maximal computational power for vanishing Lyapunov exponent~\cite{RNNchaos3}. Moreover, as mentioned earlier, state-of-the-art deep neural networks are characterized by connectivity matrices with power-law spectral densities described by universality classes.  

In biology, dynamical criticality is believed to be evolutionarily favorable. Brain activity operates in a critical state between stability, characterized by damped response to stimuli, and
epileptic chaos, characterized by exponentially sensitive response~\cite{brain 1,brain 2,brain 3,brain 4}. Examples of dynamical criticality abound in the natural world~\cite{living}, from gene expression~\cite{gene expression}, to cell growth~\cite{cell growth}, to flock dynamics~\cite{flock obs,flock dynamics}. It has been conjectured that this was evolutionarily selected because dynamical criticality offers complex systems an ideal trade-off between robust, reproducible response and flexibility of adaptation to a changing environment. 

Similarly our mechanism can be interpreted as natural selection of watchers, by analogy with optimal foraging theory~\cite{OFT}, a rich subject in evolutionary ecology. A watcher in a given region can be thought of as a forager searching for scarce resources (hospitable vacua). The network topology and transition rates in that region (foraging site) define the foraging strategy of the watcher, while the probability of tunneling into terminal vacua encodes the risk of predation. Watchers with inefficient foraging strategies die before finding resources, while those with optimal foraging strategy find resources early on. This can be understood as natural selection of watchers, with `fitness' defined as short average MFPT relative to the characteristic decay time into terminals. Furthermore, the optimal search strategy corresponds to critical dynamics.\footnote{The idea that natural selection can play a role in cosmology was proposed long ago by Smolin~\cite{Smolin:1990us,Smolin:1994vb,Smolin:1995ug,Smolin:1997pe}. His scenario postulates that space-time regions inside black holes form new universes. Whenever this happens, theory parameters undergo small random mutations which eventually favor universes that maximize black hole production. Another example of natural selection in the landscape was proposed in~\cite{Brown:2013fba}.}

\subsection{Phenomenological implications}

The search optimization mechanism makes concrete phenomenological predictions for vacua in optimal regions, described in Sec.~\ref{pheno}.
We should stress emphatically that these predictions do not rely on anthropic reasoning --- they follow generally from optimality.  

First and foremost, the critical transition rate~\eqref{tau crit intro} implies an optimal lifetime for our vacuum, 
\be
\tau \sim \frac{M_{\rm Pl}^2}{H_0^3} \sim 10^{130}~{\rm years}\,.
\label{tdecay pred intro}
\ee
This explains the metastability of the electroweak vacuum. Thus, given the observed vacuum energy~$\sim H_0^2$, we predict the 
optimal lifetime for our vacuum, which in turn is sensitive to electroweak physics (in particular the Higgs and top quark masses). 
Quantitatively, the predicted lifetime agrees with~\eqref{tdecay obs intro} to within~$\gsim\; 2\sigma$. Closer agreement 
can be achieved if the top quark is slightly heavier, $m_{\rm t} \simeq 174.5~{\rm GeV}$, which can be viewed as a prediction.
This assumes, of course, that the SM is valid all the way to the Planck scale. New physics at intermediate scales, such as
right-handed neutrinos with mass of $10^{13}$-$10^{14}~{\rm GeV}$~\cite{EliasMiro:2011aa}, can bring the expected
lifetime of our vacuum closer to the optimal prediction~\eqref{tdecay pred intro}.

In general, our mechanism selects regions with efficient transition rates, particularly for low-lying vacua, and thus
gives a {\it raison d'\^{e}tre} for the inferred Higgs metastability. New physics at scales below the SM instability scale,~$\sim 10^{10}~{\rm GeV}$, that affect the metastability of our vacuum, are disfavored by our mechanism. A prime example is low-scale SUSY~\cite{Giudice:2011cg}. As argued in Sec.~\ref{pheno}, a natural prediction of our mechanism is that the SUSY-breaking
scale is $\gsim\;10^{10}~{\rm GeV}$. Thus optimal regions of the landscape are characterized by very high-scale SUSY breaking,
which is consistent with the absence of low-scale SUSY at the LHC.

\section{Random Walks on Complex Networks}
\label{RW complex nets}

Consider a finite region of the landscape, comprised of $N\gg 1$ metastable dS vacua with potential energy $V_j > 0$, where $j = 1,\ldots, N$. 
We assume that $N$ is large enough that the region includes some number of hospitable target vacua. Statistically the $V_j$'s are assumed to be drawn from some fixed probability distribution, which, at least for $V_j$ much smaller than the fundamental scale, we may assume to be a uniform distribution \cite{Weinberg:2000qm}. (More precisely, \cite{Weinberg:2000qm} argues for a uniform probability distribution in the narrow range of ``anthropically allowed'' vacua.) The region also contains AdS/Minkowski terminal vacua, which act as
absorbing nodes or probability sinks.\footnote{In a separate paper we will consider an alternative, albeit more speculative possibility, namely that collapsing AdS regions can sometimes bounce and avoid big crunch singularities, as considered by~\cite{Garriga:2012bc,Garriga:2013cix}.} We will not explicitly keep track of terminals --- their only role is to introduce a rate of ``death" for watchers. In other words, the occupational probabilities $f_i(t)$ defined below are conditional on a watcher not having yet transitioned to a terminal vacuum and died.  

The region is modeled as a network (or graph) of $N$ sites/nodes representing the different vacua (Fig.~\ref{region intro}),
with links denote the relevant transition rates. For the time being, the region is approximated as a closed system, ignoring the exchange
of probability with its surroundings. We will relax this assumption in Sec.~\ref{rec vs tran}. Following the seminal papers by Garriga, Vilenkin and collaborators~\cite{Garriga:1997ef,Garriga:2005av}, cosmological evolution on the landscape can be modeled as a simple Markov process. Let $f_i(\tau_i)$ denote the probability that the watcher occupies vacuum $i$, as a function of the local proper time $\tau_i$. Equivalently, $f_i$ is the fraction of total comoving volume occupied by vacuum $i$. The occupation probability satisfies a forward master equation:\footnote{The master equation is linear in the transition rate because it ignores the effect of bubble collisions. More generally, bubble collisions can be encoded perturbatively as corrections that are non-linear in the transition rates~\cite{Salem:2012wa}.} 
\be
{\rm d}f_i = \sum_j {\rm d}\tau_j \kappa^{\text{proper}}_{ij}f_j - {\rm d}\tau_i \sum_r \kappa^{\text{proper}}_{ri} f_i\,,
\label{master0}
\ee
where $\kappa^{\text{proper}}_{ij}$ is the $j \rightarrow i$ transition rate. We define the time variable $t$ as~\cite{Garriga:2005av,Vanchurin:2006qp}:  
\be
{\rm d}\tau_i = {\cal N}_i {\rm d}t\,,
\label{proper global} 
\ee
where ${\cal N}_i$ is a lapse function. In a discrete setting, we can think of $t$ as a uniform discrete
counter for transitions. In our analysis we will remain agnostic about the choice of time. Crucially later on we will focus on random walk statistics,
specifically recurrence {\it vs} transience, that are time-reparametrization invariant. 

In terms of time $t$, the master equation~\eqref{master0} takes the form
\be
\frac{{\rm d}f_{i}}{{\rm d}t} = \sum_j M_{ij}f_{j} \,,
\label{master}
\ee
where $M_{ij}$ is the transition matrix:
\be
M_{ij} \equiv \kappa_{ij} - \delta_{ij} \sum_r \kappa_{rj} \,,
\label{M}
\ee
and 
\be
\kappa_{ij} = {\cal N}_j \kappa^{\text{proper}}_{ij} 
\label{lapse}
\ee
is the $j \rightarrow i$ transition rate as per unit time $t$. Henceforth we will work with $t$ for convenience; we will change back to proper time at the end by using~\eqref{lapse}. By construction, the sum of any column of $M$ vanishes, $\sum_i M_{ij} = 0$, which implies
\be
\sum_{i=1}^N f_{i} = 1\,.
\label{sum f_i}
\ee
Thus probability is conserved, since by assumption the watcher has not yet died. 

It is convenient to expand $f_i(t)$ in terms of Green's functions $P_{ki}(t)$:
\be
f_i(t) = \sum_j P_{ij}(t) f_j(0)\,.
\label{f_i t}
\ee
Here, $P_{ki}(t)$ is the $N\times N$ occupational probability matrix that the system is in state $k$ at time $t$, starting from state $i$ at $t = 0$.
It satisfies~\eqref{master} 
\be
\frac{{\rm d}P_{ki}}{{\rm d}t} = \sum_j M_{kj}P_{ji} \,,
\label{forward_master}
\ee
with initial condition $P_{ki}(0) = \delta_{ki}$. More succinctly, in matrix notation,
\be
\frac{{\rm d}P}{{\rm d}t} = MP\,;\qquad P(0) = \mathbb{1} \,.
\ee
The solution is
\be
P(t) = {\rm e}^{Mt}\,.
\label{P soln}
\ee

\subsection{Spectrum of $M$ and statistical equilibrium}
\label{Mspectrum}

The transition matrix $M$ satisfies the following properties: $i)$ its off-diagonal elements are positive semi-definite, $M_{ij} = \kappa_{ij} \geq 0$, for $i\neq j$; and $ii)$ the sum of any column vanishes, $\sum_i M_{ij} = 0$. We further assume that it is {\it irreducible}, {\it i.e.}, there exists a sequence of transitions $i\rightarrow \ldots\rightarrow  j$ connecting any $(i,j)$ pair. 

With these assumptions, one can show by invoking Perron-Frobenius' theorem that the maximal eigenvalue of $M$ is non-degenerate and vanishes,
\be
\lambda_1 = 0\,,
\label{lambda0}
\ee
while all other eigenvalues are strictly negative, $0 > \lambda_2 \geq \ldots \geq \lambda_{N}$~\cite{Garriga:2005av}. The zero-mode $v^{(1)}_M$ satisfies
\be
M v^{(1)}_M = 0\,.
\label{Mzero}
\ee
Per~\eqref{master}, this state sets the stationary distribution,
\be
f^\infty_i  \equiv f_i(t\rightarrow \infty) = v^{(1)}_{M\,i}\,.
\label{f_i 0M}
\ee
The characteristic time required to reach this distribution, known as the mixing time, is set by the
smallest (in magnitude) non-zero eigenvalue:
\be
t_{\rm mix} = |\lambda_2|^{-1}\,.
\label{tmix}
\ee

\subsection{Transitions rates}
\label{transitions section}

While transitions between any two vacua are possible quantum mechanically, whether or not they are connected by an instanton~\cite{Brown:2011ry},
in practice the decay rate out of a given vacuum $j$ is dominated by transitions to a small number of destination vacua. These are shown in
Fig.~\ref{region intro} as links between nodes. The set of links specifies the topology of the network. 

For concreteness we assume such transitions occur through Coleman-De Luccia (CDL) tunneling, in which case the $j\rightarrow i$ transition rate
is of the weighted form
\be
\kappa_{ij} = \frac{A_{ij}}{w_j}\,;\qquad A_{ij} = A_{ji}\,.
\label{kappa general}
\ee
The adjacency matrix $A_{ij}$ is given by, in the saddle-point approximation, 
\be
A_{ij} = \left(\Lambda^4 {\rm e}^{-S_{\rm bounce}}\right)_{ij}  \,,
\ee
where $S_{\rm bounce}$ is the bounce action, and $\Lambda^4$ is a characteristic scale set by the fluctuation
determinant. The weight $w_j$ of each node is 
\be
w_j = H_j^3 {\cal N}_j^{-1} {\rm e}^{S_j}\,.
\label{weights}
\ee
The factor of $H_j^3$ arises from the volume of a Hubble patch of de Sitter false vacuum, where $H_j^2 = V_j/3M_{\rm Pl}^2$. The lapse
function converts from a rate per unit false-vacuum proper time to a rate per unit time~$t$, via~\eqref{lapse}. The de Sitter entropy of the false
vacuum, $S_j \equiv \frac{48\pi^2M_{\rm Pl}^4}{V_j}$, appears in the exponential, hence low-lying vacua are exponentially weighted. 

Since $A_{ij}$ is symmetric~\cite{Lee:1987qc}, the rate satisfies a form of detailed balance,
\be
\frac{\kappa_{ji}}{\kappa_{ij}} = \frac{w_j}{w_i}  = \frac{H_j^3}{H_i^3}\frac{{\cal N}_i}{{\cal N}_j} {\rm e}^{-\left(S_i-S_j\right)}\,.
\label{detailed balance}
\ee
That is, the forward {\it vs} backward rate is fixed by the relative weight of the nodes.

\subsection{Weighted networks}
\label{gen simple kappa}

The simple weighted form~\eqref{kappa general} allows for a number of analytical results~\cite{Zhang:2013}. To begin with, despite not being symmetric, $M$ has real eigenvalues, and its eigenvectors form a complete basis. To see this, note that in matrix notation
\be
M = ZW^{-1}\,,
\label{M simple}
\ee
where 
\be
Z_{ij} \equiv A_{ij} -  \delta_{ij} \sum_r A_{rj}\,;\qquad W = {\rm diag}(w_1,w_2,\ldots,w_N)\,.
\ee
Then, define the auxiliary matrix
\be
\Sigma \equiv W^{-1/2}ZW^{-1/2} = W^{-1/2} M W^{1/2}\,.
\label{Gamma M}
\ee
This matrix is symmetric, hence its eigenvalues are real, and its eigenvectors form an orthonormal and complete basis. Furthermore, because $\Sigma$ and $M$ are related by a similarity transformation, they have identical spectra. Their eigenvectors are also simply related. Denoting the eigenvectors of $\Sigma$ by $v^{(\ell)}$, $\ell = 1,\ldots, N$, with eigenvalues $\lambda_\ell$, those of $M$ are given by $W^{1/2} v^{(\ell)}$ with the same eigenvalues. In particular, the zero-modes are related
by $v^{(1)}_M = W^{1/2}v^{(1)}$.    

In terms of $\Sigma$, the solution~\eqref{P soln} for $P$ becomes
\be
P(t)  = W^{1/2} {\rm e}^{\Sigma t} W^{-1/2} = W^{1/2} \left( v^{(1)} v^{(1)\top} + \sum_{\ell=2}^{N} {\rm e}^{\lambda_\ell t} \, v^{(\ell)} v^{(\ell)\top} \right) W^{-1/2}\,.
\label{P eigen}
\ee
As $t\rightarrow\infty$, only the zero-mode contributes:
\be
P(t\rightarrow \infty) =  W^{1/2} v^{(1)} v^{(1)\top}W^{-1/2} \,.
\ee
It is straightforward to derive an explicit expression for the zero-mode. First note that the vector with unit entries, $\vec{e} \equiv (1,1,\ldots,1)$, is a zero-eigenvector of $Z$, that is,
$Z \vec{e} = 0$. It follows from~\eqref{Gamma M} that $\Sigma \left(W^{1/2} \vec{e}\right) = 0$, hence $v^{(1)} \sim W^{1/2} \vec{e}$. Normalizing, we obtain
\be
 v^{(1)}= \left(\sqrt{\frac{w_1}{w}},\sqrt{\frac{w_2}{w}},\ldots,\sqrt{\frac{w_N}{w}}\right)\,;\qquad w \equiv \sum_i^N w_i\,.
\label{zero mode}
\ee
The corresponding zero-mode of $M$ is $v^{(1)}_M = W^{1/2}v^{(1)}$. Normalizing again, we find   
\be
v^{(1)}_M =  \left(\frac{w_1}{w},\frac{w_2}{w},\ldots,\frac{w_N}{w}\right)\,.
\ee
In other words, $P_{ij} (t\rightarrow \infty) = \frac{w_i}{w} e_j$, which gives the equilibrium distribution:
\be
f^\infty_i  = \frac{w_i}{w} \,.
\label{equi}
\ee
Explicitly, using~\eqref{weights}, we have~\cite{Garriga:2005av,Vanchurin:2006qp}:
\be
f_i^\infty = \frac{H_i^3 {\cal N}_i^{-1} {\rm e}^{S_i}}{\sum_j H_j^3 {\cal N}_j^{-1}{\rm e}^{S_j}}\,.
\label{local equi}
\ee
Famously this distribution depends on the choice of time, which is one aspect of the measure problem. 
Nevertheless the dependence enters as a mild power-law correction --- the distribution exponentially
favors low-lying vacua, {\it i.e.}, those with highest dS entropy. 

\subsection{Downward approximation}

In our analysis we will need the complete spectrum $\lambda_\ell$ of the transition matrix. 
This exact spectrum requires diagonalizing an $N\times N$ matrix, which is computationally intensive for large~$N$.
A useful and intuitive approach is the ``downward" approximation~\cite{SchwartzPerlov:2006hi,Olum:2007yk},
which neglects upward transitions. To see how this is justified, recall the detailed balance condition~\eqref{detailed balance}.
Importantly, this relation only depends on the false and true vacuum potential energy. It does not depend on the potential barrier, nor does it not rely on the thin-wall approximation. The upshot is that the rate for ``upward" CDL tunneling between two vacua is exponentially suppressed relative to the downward rate. The downward approximation amounts to neglecting all upward transitions to leading order.

It is convenient to label dS vacua in order of increasing potential energy, $0< V_1 \leq V_2  \leq \ldots \leq V_N$. In this case the transition matrix becomes upper-triangular,
hence its eigenvalues are simply given by the diagonal elements:
\be
\lambda_j \simeq M_{jj} = - \kappa_j \qquad (\text{downward approximation})\,,
\label{downward lambdas}
\ee
where $\kappa_j \equiv \sum_r \kappa_{rj}$ is the total rate out of vacuum $j$. The lower-triangular matrix encodes upward transitions and can be included as a perturbative correction~\cite{SchwartzPerlov:2006hi,Olum:2007yk}. To leading order in the approximation, the lowest-energy vacuum, $V_1$, is necessarily stable, {\it i.e.}, $\kappa_1 = 0$.
This sets the zero-mode. There may be other vacua in the region whose only allowed transitions involve up-tunneling. Such vacua would also become approximately stable
to leading order in the downward approximation. In other words, their decay rate is subleading in downward perturbation theory. 

\subsection{Initial conditions}
\label{init cond}

Since all but a measure zero of watchers will sample a finite number of bubbles before entering a terminal vacuum, our approach, like other local measures, is sensitive to initial conditions. 
A possible argument, originally advocated in the context of the causal diamond measure~\cite{Bousso:2006ev,Bousso:2009dm}, is that the theory of initial conditions is in principle distinct from the measure problem and should be provided by the theory of quantum gravity. In our case, while we cannot claim complete insensitivity, we would nevertheless like to argue that our scenario is largely insensitive to initial conditions under certain assumptions. 

Specifically, we suppose that our watcher starts out in a highly-perched initial dS vacuum in the bulk of moduli space, with Planckian or string-scale energy density. The decay rate of this incipient vacuum is also assumed to be near the fundamental scale, but long-lived enough to trigger eternal inflation. We further assume that the initial vacuum is surrounded by similarly Planckian/short-lived vacua, such that diffusion from the initial node proceeds rapidly, within a few Planck/string times. 

As mentioned in the Introduction, our ensemble of watchers is generated by following all future ``decohered" classical histories of the watcher~\cite{Nomura:2011dt}. This gives an ensemble of world-lines weighted unambiguously by the quantum-mechanical branching ratios. Given our assumptions, the different watchers\footnote{To simplify the discussion, we now refer to the different branches of the watcher simply as ``different" watchers.} will therefore rapidly disperse and reach different regions near the boundary of moduli space, where low-lying, long-lived vacua presumably reside~\cite{Polchinski:2006gy}. These different, distant regions correspond to our ensemble of foraging sites. Thus, within a short time, watchers access a large number of regions (including optimal ones), and these constitute our ensemble of regions. More precisely, if we imagined coarse-graining the Markov process over large regions of the landscape, the required time would be the sampling time for the coarse-grained random walk. While the subject of coarse-grained (or ``lumped") Markov processes has been extensively studied (see~\cite{kemeny}), without detailed knowledge of the string landscape it would seem futile to estimate the coarse-grained sampling time. It is reasonable, however, that it ought to be vastly shorter than the (fine-grained) exponentially-long global mixing time for the landscape.

Once a watcher reaches a given region of the landscape, the average time needed to access any target node in the region randomly-picked according to the stationary distribution is given by the average MFPT, otherwise known as Kemeny's constant~\cite{kemeny}. As we will review in Sec.~\ref{MFPT section}, a remarkable (and counterintuitive) property of Kemeny's constant is that it is
independent of the starting node. In other words, {\it the average time required to reach any node in the region is independent of the watcher's initial conditions in that region.} In optimal regions, in particular, the relative occupational probabilities of the watcher will rapidly reach their (local) stationary distribution~\eqref{local equi}, thanks to the healthy transition rates between vacua in such regions. Thus, as long as there is sufficient time for optimal regions to be accessed, the Markovian dynamics will efficiently drive the occupational probabilities within these regions towards local equilibrium.

\section{First-Passage Processes}
\label{MFPT section}

A popular measure of random search efficiency is the mean first-passage time (MFPT)~\cite{MFPT book}, defined as the average time for a random walk to
reach a target for the first time. The MFPT has been used in a wide range of contexts, {\it e.g.},~\cite{complex network PRL,MFPT1,MFPT2,MFPT3}, though to our knowledge
this is the first application in the context of string landscape dynamics.

Our starting point is the ``survival" probability $S_{k i}(t)$, defined as the probability that a random walker starting at node
$i$ at $t = 0$ has {\it not} reached site $k$ by time $t$. An important related concept is the {\it first-passage probability density}, $F_{k i}(t)$, which represents 
the probability density that node $k$ is being visited for the first time at time $t$, given that the walker started at $i$ at $t=0$.
It is related to the survival probability $S_{k i}(t)$ by
\be
S_{k i}(t) = 1 - \int_0^t {\rm d}t' \,F_{k i}(t')\,.
\label{SF reln}
\ee
It follows that $F_{k i}(t) = - {\rm d}S_{k i}/{\rm d}t$. For a connected network with finite $N$ number of nodes, the system is guaranteed to hit any target site given sufficient time: 
$S_{k i}(t) \rightarrow 0$, equivalently $F_{ki}(t) \rightarrow 1$, as $t\rightarrow \infty$. This is intimately related to
the recurrence property of random walks on such networks, which will be the subject of Sec.~\ref{rec vs tran}. Note that we defined $F_{ki}$ as the probability density that one \textit{starts} at node $i$ and ends at node $k$ at time $t$. One can alternatively consider a \textit{conditional first passage probability} $F_{ki}^{(0)}(t)$ defined as the probability density that one \textit{leaves} the node $i$ at $t= 0$ and ends at $k$ at time $t$. The two quantities are simply related with each other. See the Appendix for further discussion.

The mean first-passage time (MFPT) is defined as the first moment of $F_{k i}(t)$:
\be
\langle t_{i\rightarrow k}\rangle = \int_0^\infty {\rm d}t\,t F_{k i}(t) =  - \frac{{\rm d}\tilde{F}_{ki}(s)}{{\rm d}s}\bigg\vert_{s = 0} \,,
\label{MFPT def}
\ee
where $\tilde{F}_{ki}(s) = \int_0^\infty {\rm d}t\, F_{ki}(t) {\rm e}^{-st}$ is the Laplace transform.
This represents the average time taken by a random walker to reach $k$, starting from $i$ at $t = 0$, averaged on all
paths connecting the two nodes. The {\it global} MFPT, otherwise known as Kemeny's constant~\cite{kemeny},
is the average time taken by a random walker to reach a target node randomly-picked according to the stationary distribution:
\be
t_{\rm MFPT}  \equiv \sum_k \langle t_{i\rightarrow k}\rangle f_k^\infty\,.
\label{average MFPT}
\ee
It is well-known that, remarkably, {\it this quantity is independent of the starting node $i$.} Another classic result of random walk theory is
that Kemeny's constant can be neatly expressed as a spectral sum over the non-zero eigenvalues of the transition matrix: 
\be
t_{\rm MFPT}  = \sum_{\ell = 2}^N \frac{1}{|\lambda_\ell|}\,.
\label{kemeny MFPT}
\ee

For completeness, we briefly review how~\eqref{kemeny MFPT} is derived for weighted random walks~\cite{Zhang:2013}.
For this purpose it is convenient to work with discretized time, defined by $t = n \Delta t$, where $n$ is an integer. A well-known and important relation
between first-passage and occupational probabilities is~\cite{MFPT book}
\be
P_{k i}(n) = \delta_{k i}\delta_{n0} +\sum_{m=0}^n \,F_{k i}(n-m) P_{kk}(m) \Delta t\,,
\label{PFt}
\ee
where we have dropped corrections of $\mathcal{O}\left((\Delta t)^2\right)$. This relation is easy to understand. Aside from the obvious initial Kronecker delta, the occupational probability at the $n^{\rm th}$ time step is given by the sum of all first-passage probabilities to $k$ at earlier times $n-m$, multiplied by the ``loop" probability $P_{kk}(m)$ that the walker started and returned to $k$ in $m$ time steps. Taking the discrete Laplace transform, defined as $\tilde{P}_{k i}(s) = \sum_{n=0}^\infty P_{k i}(n) {\rm e}^{-s n\Delta t} \Delta t$, gives 
\be
\tilde{P}_{k i}(s) = \delta_{k i}\Delta t + \tilde{F}_{k i}(s)\tilde{P}_{kk}(s)\,.
\label{PFs}
\ee
The first-passage probability for $i\neq k$ (also known as the `first-hitting' probability) is thus
\be
\tilde{F}_{k i}(s) =  \frac{\tilde{P}_{k i}(s)}{\tilde{P}_{kk}(s)}\,.
\label{first hit 1}
\ee
Using~\eqref{P eigen}, the Laplace transform of the occupational probability is given by
\be
\tilde{P}_{k i}(s) = \frac{1}{s} f_k^\infty + \sum_{\ell=2}^{N} \frac{1}{s-\lambda_\ell} \, v^{(\ell)}_k v^{(\ell)}_i  \sqrt{\frac{f_k^\infty}{f_i^\infty}}\,.
\label{P transformed}
\ee
Substituting into~\eqref{first hit 1}, we obtain
\be
\tilde{F}_{k i}(s)  = \frac{f_k^\infty + \sum_{\ell=2}^{N} \frac{s}{s-\lambda_\ell} \, v^{(\ell)}_k v^{(\ell)}_i  \sqrt{\frac{f_k^\infty}{f_i^\infty}}}{f_k^\infty + \sum_{\ell=2}^{N} \frac{s}{s-\lambda_\ell} \, v^{(\ell)\,2}_k} \,.
\ee
Differentiating with respect to $s$ and setting $s = 0$ gives the MFPT~\eqref{MFPT def} from $i$ to $k$: 
\be
\langle t_{i\rightarrow k}\rangle = - \frac{{\rm d}\tilde{F}_{ki}(s)}{{\rm d}s}\bigg\vert_{s = 0} = \frac{1}{f_k^\infty} \sum_{\ell=2}^{N} \frac{1}{|\lambda_\ell|} \left(v^{(\ell)\,2}_k - v^{(\ell)}_k v^{(\ell)}_i  \sqrt{\frac{f_k^\infty}{f_i^\infty}}\right)\,.
\ee
Using the orthonormality conditions $\sum_k v^{(\ell)\,2}_k = 1$ and $\sum_k v^{(\ell\neq 1)}_k\sqrt{f_k^\infty} \sim \sum_k v^{(\ell\neq 1)}_k v^{(1)}_k = 0$ , the average MFPT~\eqref{average MFPT} is readily obtained 
\be
t_{\rm MFPT}  = \sum_k \langle t_{i\rightarrow k}\rangle f_k^\infty = \sum_{\ell = 2}^N \frac{1}{|\lambda_\ell|}\,.
\ee
This establishes~\eqref{kemeny MFPT}.

For our purposes it will be convenient to work with a dimensionless average MFPT, defined as
\be
{\cal T}_{\rm MFPT} \equiv \frac{t_{\rm MFPT}}{\Delta t} = \sum_{\ell = 2}^N \frac{1}{|\lambda_\ell|\Delta t}
\label{dimensionless MFPT}
\ee
This quantity is time-reparametrization invariant and counts the number of time steps taken by a random walker.
It is most intuitive and easiest to calculate in the downward approximation, where the eigenvalues are given by~\eqref{downward lambdas}:
\be
{\cal T}_{\rm MFPT} \simeq \sum_{j = 2}^N \frac{1}{\kappa_j\Delta t}\,.
\label{MFPT down}
\ee
Thus the average MFPT is equal to the sum over the characteristic number of waiting time steps, $\left(\kappa_i\Delta t\right)^{-1}$, for all metastable vacua in the region. To obtain this result, we have followed the downward convention of labeling the vacua in order of increasing potential energy, as described in Sec.~\ref{transitions section}. In particular, the sum excludes the lowest-energy vacuum, $j = 1$, which is necessarily stable in the downward approximation. Importantly, since $\kappa_i$ is the decay rate per unit time $t$, the dimensionless MFPT is manifestly time-reparametrization invariant, {\it i.e.}, independent of the choice of time variable~$t$. 

A typical region in the landscape will, generically, include multiple metastable vacua whose only possible transitions involve up-tunneling. This is sketched in Fig.~\ref{frustrated region}. 
To leading order in the downward approximation, such vacua become absolutely stable ($\kappa_j \sim 0$), implying a divergent average MFPT~\eqref{MFPT intro}.
To get a finite result, one must work to next order in the downward approximation, but it is clear that the result will be an exponentially-long MFPT.
This corresponds to frustrated dynamics. A watcher in such a region has an inefficient foraging strategy and is overwhelmingly likely to tunnel to an AdS
(or Minkowski) terminal and die well before accessing a hospitable vacuum.

In the vastness of the landscape, there ought to be rare but optimal regions, whose vacua all have allowed downward transitions (except of course for the lowest-lying vacuum). 
Thus their topography is that of a broad energy valley or funnel, as shown in Fig.~\ref{optimal region}. This is akin to the {\it principle of minimal frustration} of protein energy landscapes~\cite{proteins1,proteins2}, where the high-energy unfolded states are connected to the lowest-energy native state by a relatively smooth funnel. Watchers in such regions have an efficient foraging strategy and find resources (hospitable vacua) before dying in a terminal vacuum. Thus our mechanism can be interpreted as natural selection of watchers, with `fitness' defined as short average MFPT relative to the characteristic decay time into terminals.

Another instance of funnel topography occurs in the loss/optimization landscape of deep neural networks. In general, finding the global minimum of the loss landscape is an~\textsf{NP}-complete problem~\cite{DNN NP}. Recently it has been shown that neural networks with poor training parameters correspond to highly non-convex landscapes, with many local minima, whereas well-trained networks have smooth, funnel-like landscapes. Relatedly, the connectivity matrix of well-trained, state-of-the-art deep neural networks have power-law spectral density~\cite{DNNpowerlaw}, well-described by heavy-tailed random matrix theory~\cite{RMT}, indicating that well-trained networks operate at criticality. 

To minimize the MFPT~\eqref{MFPT down}, downward transitions should be as fast as possible. However, as we will see in Sec.~\ref{rec vs tran},
transition rates are bounded from above by demanding efficient sampling of local regions, since high transition rates can cause a watcher to escape an efficient region. These considerations will lead us to define a dimensionless mean residency time. We will find that the mean residency time grows polynomially in $D=\log N$, specifically as
$\log N$ in the optimal case. Incidentally this does not contradict the~\textsf{NP}-hard complexity classification~\cite{Denef:2006ad}. \textsf{NP}-hardness is a worst-case assessment, and as such does not preclude the existence of polynomial-time solutions for special, finely-tuned instances of the problem.

As a toy example of how such a phase transition in complexity can occur in special instances of a problem, let us consider a landscape which has the topology of a tree~\cite{compPT2}. By this we mean that every vacuum in the landscape is connected to $b$ other vacua, so that there is one vacuum at depth zero, $b$ vacua at depth one, $b^2$ vacua at depth two and so on. By connectedness, we mean that each vacuum has a finite rate $\kappa_{d\to d+1}$ of transition to one of its $b$ descendants, where we have assumed for simplicity that the transition rate only depends on the depth in the tree. The total rate of transition out of a vacuum at depth $d$ is $\kappa_d = b\kappa_{d\to d+1}$. The average MFPT is therefore given by
\be
t_{\text{MFPT} }  = \sum_{d=0}^{D-1} \frac{b^{d}}{b\kappa_{d\to d+1}}\, .
\ee
As a further simplification, let us assume that $\kappa_{d \to d+1} \sim z^{-d}$ for some parameter $z$. This then gives
\be
t_{\text{MFPT} } = \frac{1 - (bz)^{D}}{b(1-bz)}\, .
\ee
Note that the MFPT is controlled by the effective parameter $p = bz$. When $p >1$, the MFPT grows exponentially with the depth, $t_{\text{MFPT}} \sim {\rm e}^{D\ln p}$; when $p = 1$, we have linear growth with depth, $t_{\text{MFPT}} \sim D$; while for $p <1 $ the MFPT is $\mathcal{O}(1)$ in the large $D$ limit.
 
\section{Recurrent and Transient Random Walks}
\label{rec vs tran}

Up to this point we have treated the regions in our ensemble as closed systems, ignoring the exchange of
probability with their surroundings. In this idealized framework, we found that the MFPT~\eqref{MFPT down}
can in principle be made arbitrarily small by dialing up transition rates. More realistically, however,
we should treat the regions as open systems, allowing the possibility that a watcher escapes
a region {\it before} hitting a target vacuum. This introduces a trade-off. Once a random walker lands in an
optimal region, the dynamics should be such that the walker efficiently explores the region, thereby
minimizing the MFPT, while at the same time minimizing the likelihood of escape before finding viable vacua. 

Instead of treating regions as open systems, which would require modeling
their environment, we focus on the more tractable (and less model-dependent) problem
of random walk dynamics in a given region in the limit of infinite volume, $N\rightarrow \infty$. To be clear, 
our model remains unchanged --- an ensemble of many regions, each with {\it finite} $N$ nodes.
However, to gain insights on the impact of their environments, we consider the behavior of random walks
in the hypothetical $N\rightarrow \infty$ limit. 

In this limit we demand that random walks be {\it recurrent}, {\it i.e.}, that random walkers will return 
to their starting point, and will do so infinitely-many times in the future, with unit probability. 
Equivalently, in recurrent walks every site will be visited with probability one. Conversely, {\it transient}
walks are such that the probability of never returning to the starting point is finite. Although not formally
equivalent to modeling regions as open systems, we advocate recurrence as a reliable and model-independent
benchmark for efficient sampling. Recurrent walks thoroughly explore any region around their starting point,
whereas transient walks tend to escape to infinity.

\subsection{Recurrence condition}
\label{rec vs tran subsec}

Recall the survival probability $S_{ki}(t)$, defined in Sec.~\ref{MFPT section} as the probability that a random walker starting at node
$i$ at $t = 0$ has not reached site $k$ by time $t$. It follows that $S_{ii}(t)$ is the probability that the walker {\it has not returned} to the starting node by time~$t$. 
The escape probability is naturally defined as the probability that the walker never returns to the starting node (see Appendix~\ref{App:FPP} for further discussion): 
\be
{\rm Escape}~{\rm probability} = \lim_{t\rightarrow \infty} S_{ii}(t) \,.
\label{escape}
\ee
A random walk is said to be {\it recurrent} if the escape probability vanishes; it is said to be {\it transient} if the
escape probability is finite. That is,
\bea
\nonumber
\lim_{t\rightarrow \infty} S_{ii}(t) &=& 0 \qquad\;\;\;\;\;\; \Longleftrightarrow \qquad {\rm recurrence} \\ 
\lim_{t\rightarrow \infty} S_{ii}(t) & = & {\rm finite} \qquad \Longleftrightarrow \qquad {\rm transience} \,.
\label{recur relation S}
\eea

This criterion can be expressed in terms of the first-passage probability density $F_{k i}(t)$, defined in~\eqref{SF reln} as the probability density that a random walk starting at
$i$ at $t = 0$ reaches $k$ for the first time at time $t$. Thus $F_{ii}(t)$ is naturally interpreted as the {\it first-return} probability density. 
Using~\eqref{SF reln}, we see that the escape probability satisfies 
\be
\lim_{t\rightarrow \infty} S_{ii}(t) = 1 - \int_0^\infty {\rm d}t \,F_{ii}(t) = 1 - \tilde{F}_{ii}(0)\,,
\label{escape reln}
\ee
where $\tilde{F}_{ii}(0) = \int_0^\infty {\rm d}t\, F_{ii}(t)$ is the probability that the random walker {\it ever} returns to $i$. It follows that
\bea
\nonumber
\tilde{F}_{ii}(0) &=& 1\qquad \Longleftrightarrow \qquad {\rm recurrence} \\ 
\tilde{F}_{ii}(0) & < & 1 \qquad \Longleftrightarrow \qquad {\rm transience} \,.
\label{recur relation}
\eea
In other words, recurrence implies that a random walker is certain to return eventually to the starting node, and will do so infinitely-many times in the future.\footnote{Paradoxically,
it possibly takes, on average, an infinite time to do so on an infinite network, as the mean first-return time $\langle t_{i\rightarrow i}\rangle = \int_0^\infty {\rm d}t\,t \,F_{ii}(t)$ can potentially diverge.}
It is worth emphasizing that $\tilde{F}_{ii}(0) = -\int_0^\infty {\rm d}t\, \frac{{\rm d}S_{ii}}{{\rm d}t}$ is time-reparametrization invariant. 
{\it Thus the recurrence criterion is independent of the choice of time variable.}

The recurrence/transience criterion~\eqref{recur relation} can be related to the occupational probability $P_{ki}(t)$. 
Setting $k = i$ in~\eqref{PFs} gives
\be
\tilde{F}_{ii}(s) = 1 - \frac{\Delta t}{\tilde{P}_{ii}(s)}\,,
\ee
where $\Delta t$ is a discrete time step. Thus the recurrence criterion~\eqref{recur relation} is equivalent to 
\be
\dT_i \equiv \lim_{s\rightarrow 0} \frac{\tilde{P}_{ii}(s)}{\Delta t} = \infty \qquad ({\rm recurrence}) \,.
\label{recur relation 2}
\ee
Averaging over all initial sites, we define 
\be
\dT \equiv \lim_{s\rightarrow 0} \left\langle \frac{\tilde{P}_{ii}(s)}{\Delta t} \right\rangle = \lim_{s\rightarrow 0} \frac{\Tr \tilde{P}(s)}{N\Delta t}  \,.
\label{PTr}
\ee
The recurrence criterion amounts to demanding that~$\dT$ diverges. The factor of~$\Delta t$ has been kept for convenience, to make manifest the fact that $\dT$ is
time-reparametrization invariant. This will play an important role in what follows. 

The trace of the occupational probability follows readily from~\eqref{P transformed},
\be
\frac{\Tr \tilde{P}(s)}{N} = \frac{1}{Ns} + \frac{1}{N} \sum_{\ell=2}^{N} \frac{1}{s-\lambda_\ell}\,,
\ee 
where we have used~\eqref{sum f_i} and the normalization condition $\sum_i v^{(\ell)\,2}_i = 1$.
At fixed $N$, the first term, which arises from the zero-mode, diverges as $s\rightarrow 0$. Hence 
random walks are always recurrent on finite networks. To study recurrence on {\it infinite} networks,
one should first let $N\rightarrow \infty$ {\it before} sending $s\rightarrow 0$. In this limit the zero-mode
contribution vanishes~\cite{Michelitsch:2017}, leaving us with
\be
\dT = \frac{1}{N}\sum_{\ell=2}^{N} \frac{1}{|\lambda_\ell|\Delta t} \,,
\label{dT MFPT}
\ee
where the limit $N\rightarrow \infty$ is understood. Note that $\dT$ can be expressed in terms of the dimensionless average MFPT~\eqref{dimensionless MFPT} as
\be
\dT = \frac{{\cal T}_{\text{MFPT}}}{N}\,.
\ee
Therefore, we may think of $\dT$ as a dimensionless mean residency time.

Once again the result is most transparent in the downward approximation, where the eigenvalues are given by~\eqref{downward lambdas},
and the MFPT reduces to~\eqref{MFPT down}:
\be
\dT \simeq \left\langle \frac{1}{\kappa_i\Delta t}\right\rangle \qquad ({\rm downward})\,.
\ee
Since $\kappa_i$ is the decay rate per unit time $t$, this makes the time-reparametrization invariance of $\dT$ manifest.
In particular, it will be convenient to express the result in terms of proper time: $\kappa_i\Delta t = \kappa_i^{\rm proper} \Delta \tau_i$. The natural proper time step is of course the Hubble time: $\Delta\tau_i = H_i^{-1} \sim \frac{M_{\rm Pl}}{\sqrt{V_i}}$. Putting everything together,
\be
\dT \simeq  \left\langle \frac{\sqrt{V_i}}{M_{\rm Pl}\kappa_i^{\rm proper}}\right\rangle = \left\langle \frac{\sqrt{V_i}}{M_{\rm Pl}} \,\tau_i\right\rangle\,,
\label{PTr proper}
\ee
where $\tau_i \equiv 1/\kappa_i^{\rm proper}$ is the (proper) lifetime of the $i^{\rm th}$ vacuum. Whether~\eqref{PTr proper} is finite or
diverges as $N\rightarrow \infty$ determines whether the random walk is transient or recurrent, respectively.\footnote{To qualify as a semiclassical vacuum, a given vacuum should at least survive for a Hubble time, {\it i.e.}, $H_i \tau_i\;\gsim\; 1$. It follows that $\dT\;\gsim\; 1$. Using~\eqref{dT MFPT}, we learn that ${\cal T}_{\rm MFPT} \;\gsim\; {\cal O}(N)$, that is, the MFPT must scale at least linearly in $N$.}

\subsection{Statistical average}

Our next task is to express~\eqref{PTr proper} as a suitable statistical average over possible realizations of the region. In general the
rate $\kappa_i$ out of a given metastable vacuum depends on its potential energy $V_i$ as well as the shape of the potential barriers that determine
the bounce solution to the different destination vacua. Denoting the ``shape" parameters for transitions out of the $i^{\rm th}$
vacuum collectively by $\theta_i$, we have
\be
\kappa_i^{\rm proper}\equiv \kappa^{\rm proper}(V_i,\theta_i)\,.
\ee
The dependence on $V_i$ comes from two sources. An immediate source is the volume factor $H_i^{-3}\sim V_i^{-3/2}$ of de Sitter false vacuum. 
A second, and less obvious factor comes from the expected suppression of possible destination vacua as $V_i\rightarrow 0$. Indeed, the assumption of
a flat probability distribution of vacuum energy and the exponential suppression for upward tunneling together imply that, statistically, lower-lying vacua
have fewer possible dS destinations than high-lying vacua. 

Let ${\cal P}(V,\theta)$ denote the underlying joint probability distribution for a given vacuum to have potential energy $V$ and bounce parameters $\theta$. 
In the large $N$ limit the dimensionless mean residency time can be approximated via the central limit theorem by a statistical average weighted by the underlying probability distribution:
\be
\dT  = \int {\rm d}V{\rm d}\theta   \frac{\sqrt{V}}{M_{\rm Pl}} \,\tau(V,\theta) \,{\cal P}(V,\theta)\,,
\label{R1}
\ee
with corrections suppressed by $1/\sqrt{N}$. For simplicity let us assume that on the string landscape the absolute height of a
vacuum and the shape of the surrounding potential barriers are uncorrelated: 
\be
{\cal P}(V,\theta) \equiv {\cal P}(V) \hat{{\cal P}}(\theta)\,.
\ee
Equation~\eqref{R1} then simplifies to
\be
\dT  = \int {\rm d}V  \frac{\sqrt{V}}{M_{\rm Pl}}\, \tau(V) \,{\cal P}(V) \,,
\ee
where 
\be
\tau(V) \equiv \int {\rm d}\theta \,\tau(V,\theta)\hat{{\cal P}}(\theta)
\ee
is the average lifetime of a vacuum with potential energy $V$. 

Not surprisingly, the recurrence criterion $\dT\rightarrow \infty$ is sensitive to the behavior of $\tau(V)$ for the lowest-lying
(and, as argued above, generically most stable) vacua. Provided the probability distribution ${\cal P}(V)$ falls off sufficiently fast for large $V$, the divergence in $\dT$ must come from the small-$V$ region of the integral. Therefore, random walks will be recurrent if the average lifetime $\tau(V)$ diverges
sufficiently fast as $V\rightarrow 0$. Specifically, assuming as before that ${\cal P}(V)$ is nearly flat for $V$ much smaller than the fundamental scale, 
the recurrence criterion is
\be
\dT  \sim \int_0 {\rm d}V  \sqrt{V} \, \tau(V) \rightarrow \infty  \qquad ({\rm recurrence}) \,.
\label{R2}
\ee
This condition will be satisfied if $\tau(V)$ diverges faster than $V^{-3/2}$ as $V \rightarrow 0$, with the critical case $\sim V^{-3/2}$ resulting in a
logarithmic divergence.

\section{Dynamical Criticality}
\label{edge of chaos}

We have derived two competing conditions: minimal dimensionless average MFPT, which as argued in Sec.~\ref{MFPT section} requires fast transition rates;
and recurrence in the $N\rightarrow \infty$ limit, which as shown in Sec.~\ref{rec vs tran} requires sufficiently long-lived vacua. Optimal regions
reach a compromise: they achieve the shortest MFPT compatible with recurrence, {\it i.e.}, the least-divergent recurrence integral~\eqref{R2}. Thus,
optimal regions lie at the critical boundary between recurrent and transient walks, characterized by an average lifetime scaling as~$V^{-3/2}$. Using
the Planck mass $M_{\rm Pl}$ to fix dimensions, since this is the only other scale at hand, we obtain
\be
\tau_{\rm crit} (V) \sim \frac{M_{\rm Pl}^5}{V^{3/2}}\sim \frac{M_{\rm Pl}^2}{H^3}~~~{\rm as}~~V \rightarrow 0\,.
\label{tau crit}
\ee
Remarkably, this is recognized as the Page time~\cite{Page:1993wv} for dS space~\cite{Danielsson:2002td,Danielsson:2003wb,Ferreira:2016hee,Ferreira:2017ogo}. In slow-roll inflation, the Page time marks the phase transition to slow-roll eternal inflation~\cite{Creminelli:2008es} and has been used to place a bound on the maximum number of e-folds that can be described semi-classically~\cite{ArkaniHamed:2007ky}. The appearance of the Page time in false-vacuum eternal inflation, and its relation to random walk criticality, is surprising. We do not yet have a compelling intuitive explanation for its occurence.  

In any case, the critical lifetime~\eqref{tau crit} implies a logarithmically-divergent integral~\eqref{R2}.\footnote{Equation~\eqref{tau crit} captures the leading divergence as $V\rightarrow 0$. We cannot constrain slowly-varying factors, such as $V^{-3/2} \log^n V$, which, in any case, give negligible corrections to the phenomenological predictions discussed in Sec.~\ref{pheno}.} Correspondingly, at finite $N$ the mean residency time~\eqref{PTr proper} diverges as $\log N$, which signals {\it dynamical criticality}. A similar non-equilibrium phase transition occurs in quenched disordered media, when the probability distribution for waiting times reaches a critical power-law~\cite{disordered media}. Thus the joint demands of sweeping exploration, defined by recurrence, and minimal oversampling, defined by minimal MFPT, select regions of the landscape that are tuned at criticality.\footnote{There is a simple analogue in the case of simple (Brownian) random walks on regular $d$-dimensional lattices. For fixed number of sites $N$, the average MFPT decreases with increasing dimension.
For instance, $t_{\rm MFPT} = N(N+1)/6$ for $d = 1$, $t_{\rm MFPT} = \pi^{-1} N\ln N + {\cal O}(N)$ for $d = 2$, and $t_{\rm MFPT} \simeq 1.52\,N+ {\cal O}(\sqrt{N})$ for $d=3$~\cite{excitons}. On the other hand, by P\'olya's theorem simple random walks are recurrent for $d \leq 2$, and transient for $d > 2$~\cite{polya}. The minimal MFPT
compatible with recurrence is achieved in $d = 2$, the critical dimension for recurrence/transience.}

The criticality of the recurrence/transience boundary can be understood in various ways. From a computational complexity standpoint, it represents a transition in the scaling of the 
mean residency time~\eqref{PTr proper} as a function of the effective dimensionality $D$ of the landscape region. (Recall that the number of vacua scales exponentially
with the landscape dimension, $N \sim {\rm e}^D$.) Recurrent walks typically result in a power-law divergent integral in~\eqref{R2}, corresponding at finite $N$ to a
recurrence measure diverging exponentially in $D$. In the critical case, the recurrence measure instead grows logarithmically in $N$,
and therefore polynomially in $D$. More precisely, the large $N$ behavior delineates regions in the landscape into three basic types:
\be
\dT  \sim \left\{\begin{array}{ccl}
\text{Exp}(D)&\cdots & \text{(recurrent but high complexity)} \\
 \text{Poly}(D)&\cdots & \text{(recurrent and low complexity)} \\
 \text{Constant}& \cdots &  \text{(low complexity but transient)}\,.
\end{array}\right.
\ee
The second line corresponds to a critical boundary between the other two phases. The phase transition we have uncovered belongs in the category of {\it computational phase transitions}, which also occur in the infinite-size limit. These have generated much activity in computer science in recent years~\cite{compPT1,compPT2}. A famous example is the phase transition in heuristic decision-tree pruning from polynomial to exponential search time at a critical value of the effective branching ratio~\cite{pruning}. Another example is the phase transition from satisfiability to unsatisfiability in the random $k$-SAT problem at a critical value of the ratio of constraints to variables~\cite{kSAT}.

The recurrence/transience boundary is also critical in that it delineates stable and unstable dynamical behavior. To see this, consider a pair of random walks starting
from the same initial site. In the recurrent regime, the walkers are certain to meet again, infinitely-many times, in the future. In the transient regime, however, they
will eventually diverge from each other, never to meet again. Thus the pair of walks can be thought of as dynamically stable in the recurrent case, and
unstable in the transient case. 

As a proof of principle, the desired critical scaling~\eqref{tau crit} can arise in the following simplistic situation. Consider a funnel-like
region of the landscape with effective dimensionality $D$. Suppose that the potential energy is described by a smooth power-law envelope $V(\rho) \sim \rho^n$,
where~$\rho$ is the distance in moduli space, punctuated by a lattice of metastable vacua that are equally spaced on average. Thus within a given
distance~$R$ there are $\sim R^D$ vacua. Furthermore, suppose that each vacuum in the funnel can transition to all lower vacua with approximately equal
rate.\footnote{This could be achieved, for instance, with `giant leaps'~\cite{Brown:2010bc}.} Then the total transition rate out of a given vacuum with potential energy $V_j \sim R_j^n$ is proportional to the number of destinations $\sim R_j^D \sim V_j^{D/n}$. This must be divided by $H^3_j \sim V_j^{3/2}$ to obtain the proper
total transition rate: $\kappa_j^{\rm proper} \sim V^{\frac{D}{n} -\frac{3}{2}}$. This gives the desired scaling~\eqref{tau crit} for $D = 3n$. 

In the natural world there is a striking relation between complexity of self-organizing systems and criticality~\cite{living}. Examples
include neuronal activity in the brain, flocking behavior of starlings, insect swarms and cell growth, to name a few. Such systems
operate in a critical state between stable and unstable dynamical regimes. Empirical observations that support the dynamical
criticality hypothesis include, for instance, the probability distribution of neuronal avalanche size being scale-free~\cite{brain 3},
and correlation functions of velocity dispersion of flocking birds displaying scale invariance~\cite{flock obs}. It has been conjectured
that dynamical criticality is evolutionarily favored because it offers an ideal compromise between robust response to external stimuli
and flexibility for adaptation to environmental changes. Similarly, our mechanism offers a dynamical explanation for why our universe is poised at criticality.  

\section{Phenomenological Implications}
\label{pheno}

The natural selection mechanism based on search optimization outlined above has concrete phenomenological implications for
our vacuum. Importantly, these specific predictions do not rely on anthropic reasoning; instead they follow immediately from
the properties of optimal regions of the landscape. 

\vspace{0.3cm}
\noindent {\bf Lifetime of our vacuum:} If our vacuum is part of an optimal region of the landscape, characterized by vacua
with critical lifetime given the Page time~\eqref{tau crit}, then we predict a lifetime of
\be
\tau \sim \frac{M_{\rm Pl}^2}{H_0^3} \sim 10^{130}~{\rm years}\,.
\label{tdecay pred}
\ee
This explains the metastability of the electroweak vacuum. As mentioned in the Introduction, what makes the Higgs metastability
particularly interesting is that it relates the observed cosmological constant, through the $H_0^{-3}$ volume factor,
and electroweak physics, in particular the Higgs and top quark masses. Thus, taking the observed vacuum
energy $\sim M_{\rm Pl}^2 H_0^2$ as given, the optimal lifetime~\eqref{tdecay pred} can be interpreted as constraining 
the Higgs and top masses to lie around the weak scale (keeping other SM parameters fixed). 

Quantitatively, the predicted lifetime agrees with the SM prediction~\eqref{tdecay obs intro} to within $\gsim\; 2\sigma$. Closer agreement 
can be achieved if the top quark is slightly heavier, $m_{\rm t} \simeq 174.5~{\rm GeV}$. This can be viewed as a prediction, assuming of course 
that the SM is valid all the way to the Planck scale. New physics at intermediate scales can reduce the tension. For instance,
adding a gauge-invariant, higher-dimensional operator $\frac{h^6}{\Lambda_{\rm NP}^2}$ can affect the predicted lifetime
if $\Lambda_{\rm NP}\;\lsim\; 10^{13}~{\rm GeV}$, assuming the central value $m_{\rm t} = 173.5~{\rm GeV}$~\cite{Andreassen:2017rzq}. 
Another possibility are sufficiently light primordial black holes which act as catalysts for the decay~\cite{Burda:2016mou}.

More generally, our mechanism selects regions with efficient transition rates, particularly for low-lying vacua, and thus
gives a {\it raison d'\^{e}tre} for the conspiracy underlying the Higgs metastability. In other words, from our point of view the inferred
metastability of the electroweak vacuum is sacred. New physics below the SM instability scale, $\sim 10^{10}~{\rm GeV}$, on the other hand, can jeopardize this
observable. Here are the implications for some candidates of BSM physics: 

\begin{itemize}

\item {\it Low-scale SUSY:} If the SUSY breaking scale is $\lsim\;10^{10}~{\rm GeV}$, this will directly impact the stability of our vacuum. There are three obvious possibilities: 1)~SUSY makes our vacuum unstable ({\it e.g.}, via decay to charge/color breaking vacua~\cite{Gunion:1987qv,Casas:1995pd,Kusenko:1996jn,Strumia:1996pr,Abel:1998wr,Chowdhury:2013dka}), which by itself is inconsistent and would therefore requires additional new physics; 2)~SUSY makes our vacuum stable, which is disfavored by our mechanism; 3)~SUSY maintains our vacuum within the metastability region. The latter possibility is logically consistent with our mechanism but would require further numerical conspiracy, above and beyond that already achieved in the SM. Therefore, barring fine-tunings, the natural implication of SUSY below $10^{10}~{\rm GeV}$ is to make our vacuum stable, which is disfavored by our mechanism. 

This expectation is borne out by an explicit calculation of~\cite{Giudice:2011cg}, which showed that if all SUSY partners have masses at the SUSY breaking scale,
then the metastability of our vacuum requires a SUSY breaking scale of $\gsim\;10^{10}~{\rm GeV}$. Thus optimal regions of the landscape are characterized by
very high-scale SUSY breaking, which is consistent with the absence of low-scale SUSY at the LHC.

\item {\it Sterile neutrinos:} Massive right-handed neutrinos, like the top quark, tend to make the vacuum less stable. Assuming three right-handed neutrinos of comparable mass, for simplicity, the impact on Higgs metastability is negligible if their mass is $\lsim\; 10^{13}~{\rm GeV}$~\cite{EliasMiro:2011aa}. On the other hand, if their mass is around $10^{13}$-$10^{14}~{\rm GeV}$, then the expected lifetime for our vacuum will be in closer agreement with the predicted optimal lifetime~\eqref{tdecay pred}.  

\item {\it QCD axion:} Consider the QCD axion as a solution to the strong CP problem. The radial part of the~$U(1)$ complex scalar is a boson and hence makes the electroweak vacuum more stable. To preserve the desired metastability, the Peccei-Quinn scale must be sufficiently high, $f_{\rm a} \;\gsim\; 10^{10}~{\rm GeV}$~\cite{Hertzberg:2012zc}.

\end{itemize}

\vspace{0.3cm}
\noindent {\bf Cosmological constant problem:} Unlike the above predictions, which depend only on optimality, the prediction for the cosmological constant
is sensitive to the choice of measure. Assuming that vacua are weighted by comoving volume, then the local equilibrium distribution is $f_i^\infty \sim {\rm e}^{48\pi^2 M_{\rm Pl}^4/V_i}$. The distribution is sharply peaked and exponentially favors the lowest-lying vacuum in the region. Thus, within an optimal region selected by the search optimization principle, we are overwhelmingly likely to find ourselves in the vacuum with the smallest potential energy $V_{\rm min}$ within that region. 

Assuming as before that the underlying probability distribution ${\cal P}(V)$ is nearly flat for $V \ll M_{\rm Pl}$, then statistically the minimum potential energy achieved in a region of $N$ vacua is
\be
V_{\rm min} \sim \frac{M_{\rm Pl}^4}{N}\,.
\ee 
This can account for the observed cosmological constant if our region contains $N \sim 10^{120}$ vacua. In particular, the predicted value of the cosmological constant is set by the region size~$N$, which is manifestly UV-insensitive. 

Because an identical argument could be made in the ``global" approach to the landscape~\cite{Polchinski:2006gy,Bousso:2010zi,Linde:2010nt}, it is important to highlight a few key differences. The global equilibrium distribution (achieved after an exponentially-long mixing time) is identical to the aforementioned local stationary distribution, thus it exponentially favors the smallest positive vacuum energy {\it anywhere in the landscape}. This cannot be considered a satisfactory solution to the cosmological constant problem, on a number of counts. (See Sec.~5 of~\cite{Denef:2006ad} for a discussion of some of these problems.) One issue is that, from a computational complexity perspective, the problem of finding the minimal positive vacuum energy globally is harder than \textsf{NP}, {\it i.e.}, a candidate solution cannot be checked in polynomial time. The most important issue, however, is phenomenological --- vacua with the smallest positive potential energies are expected to be nearly supersymmetric, with vacuum energy~$\sim m_{3/2}^4$, thus the global measure would seem to
overwhelmingly predict a tiny SUSY breaking scale. 

These problems do not afflict the ``local" approach. The optimal regions selected by our mechanism naturally have a very high SUSY breaking
scale, as already mentioned. This addresses Banks' interpretation of the cosmological constant problem, namely ``given the value of the cosmological constant,
why is SUSY breaking so large?"~\cite{Banks:2000fe}. By construction the optimal regions comprise a finite number of dS vacua, and all one needs
to explain the observed cosmological constant is for one of these regions to have $N\sim 10^{120}$ vacua. 

\section{Conclusions}

It is striking that a number of major theoretical puzzles in fundamental physics can be interpreted as problems of near criticality. The weak hierarchy problem translates to the Higgs field being close to the phase transition between broken and unbroken electroweak symmetry~\cite{Giudice:2006sn}. The cosmological constant problem is the statement that our universe is nearly flat, Minkowski space, which in turn can be thought of as a quantum critical point between dS and AdS space-times. The approximate scale invariance of primordial density perturbations suggests near criticality in the early universe. This is realized explicitly through a period of approximate dS expansion, during which the inflating universe is conformally invariant.  

Yet another indication of criticality is the inferred metastability of the electroweak vacuum~\cite{Buttazzo:2013uya}.
The Higgs metastability, which is the result of a delicate numerical conspiracy, crucially relies on the absence of new physics above the weak scale. As such it
leaves no room for indifference. Either one concedes that SM metastability is purely a numerical coincidence, as implicitly assumed, for instance,
in all BSM theories with low-scale SUSY. Or it is a real phenomenon in want of an explanation, in which case naturalness cannot be the
answer to the weak hierarchy problem. 

In our mind, the near criticality of our universe strongly suggests a statistical physics origin. A natural arena for the statistical physics
of universes is the landscape of string theory, together with the mechanism of eternal inflation for populating vacua. In this
paper we presented a dynamical selection mechanism, based on search optimization, which favors vacua in regions of the landscape
where random walk dynamics are at criticality. This suggests a connection between the near criticality of our universe and non-equilibrium
critical phenomena in landscape dynamics.

Instead of concentrating on late-time stationary probability distributions, in this work we focused on the approach to equilibrium. The underlying assumption
is that cosmological evolution on the multiverse has occurred for a time much shorter than the exponentially-long global mixing time of the
landscape~\cite{Denef:2017cxt}. Accessing hospitable vacua then becomes a race among watchers random walking on the landscape. 
The ``winners" are watchers that land in optimal regions of the landscape where the search algorithm is efficient. 

We defined optimality by two competing requirements. The first requirement is search efficiency, quantified by the average MFPT, which requires fast transition rates.
The average MFPT is the average time taken by a watcher to reach a target vacuum, picked at random according to the stationary probability distribution. We showed that
the average MFPT is minimized for hospitable vacua lying at the bottom of funnel-like regions, akin to the smooth folding funnels of naturally-occurring
proteins and the convex loss functions of well-trained deep neural networks. The second requirement is sweeping exploration, which requires recurrence in
the infinite-volume limit. Optimal regions reach a compromise by lying at the critical boundary between recurrence and transience, thereby achieving dynamical criticality.
In other words, they have the shortest MFPT compatible with recurrence. Crucially, unlike other statistical measures on the landscape, our optimality criterion
(and the critical boundary it defines) is time-reparametrization invariant. 

Remarkably, vacua in optimal regions have lifetimes of order the dS Page time, $\tau_{\rm crit} \sim M_{\rm Pl}^2/H^3$.
The emergence of the Page time in relation with critical dynamics on the landscape is surprising to us, and we do not yet
have an intuitive explanation. In any case, for our vacuum this implies a predicted lifetime of $M_{\rm Pl}^2/H_0^3 \sim
10^{130}~{\rm years}$, which agrees with the SM prediction~\eqref{tdecay obs intro} to within $\gsim\; 2\sigma$. Closer
agreement can be achieved if the top quark is slightly heavier, $m_{\rm t} \simeq 174.5~{\rm GeV}$, or, more plausibly, 
with new physics at intermediate scales, such as $\sim 10^{13}~{\rm GeV}$ right-handed neutrinos. Barring fine-tunings, 
our mechanism predicts that the SUSY breaking scale should be high, $\gsim\; 10^{10}~{\rm GeV}$. Evidence of SUSY
at the LHC would likely rule out our scenario. 

Philosophically our proposal is diametrically opposite to the standard approach to the landscape. The principle of mediocrity amounts to
the statement that our vacuum should be typical among all hospitable vacua on the landscape. Instead, the hospitable vacua favored
by our mechanism are exceptional, in that they live in fine-tuned neighborhoods of the landscape. But they are exceptional for a
purpose --- to be easily accessed early on. Our dynamical mechanism can be interpreted as natural selection of watchers foraging different regions of the landscape. Watchers with inefficient
foraging strategies are likely to enter a terminal vacuum and die before finding a hospitable vacuum, while those with optimal foraging strategy are able
to find hospitable vacuum early on. 

Complex self-organized systems poised at criticality are ubiquitous in the natural world. This has motivated the conjecture
that dynamical criticality is favored evolutionarily because it offers an ideal trade-off between robust, reproducible response and
flexibility of adaptation to a changing environment. Furthermore, it has been argued that computational capabilities
are maximized at criticality. Similarly, our natural selection mechanism favors vacua at criticality.

The selection mechanism presented here offers many new avenues of investigation. In this paper we have treated AdS/Minkowski vacua conservatively as terminal,
acting as absorbing nodes or probability sinks. In a separate paper we will consider the more speculative possibility that collapsing AdS regions can sometimes
bounce and avoid big crunch singularities, as considered in~\cite{Garriga:2012bc,Garriga:2013cix}. See also~\cite{Nomura:2011dt,Piao:2004me,Piao:2009ku,Johnson:2011aa,Lehners:2012wz}.
The high energy density reached at the bounce would allow the space-time region to transition classically to possibly distant dS vacua. Thus AdS vacua could then be interpreted
as mediators of new, non-local transitions between dS vacua. We will argue that AdS-mediated transitions are analogous to L\'evy flights, which have been shown to improve search efficiency~\cite{levy classic,intermittent,intermittent2,intermittent3,DiPatti:2015}. Indeed, inspired by the saltatory behavior of foraging animals~\cite{levy animals,levy animals 2,sharks1,sharks2,humans}, it has been conjectured that optimal search strategies to find sparse, randomly-located targets on a network combines local (Brownian) moves with non-local (L\'evy) relocations. A classic example in network science is Google's PageRank algorithm~\cite{pagerank,pagerank review}.

The most interesting implications may be for string phenomenology and model-building. String vacua with realistic particle physics are usually considered
in isolation, without consideration for their accessibility and the topography of the surrounding landscape region. In the context of flux compactifications, it
has been argued that realistic vacua with small $\theta_{\rm QCD}$ (either because of a light axion or because of spontaneous breaking of strong CP) or
sufficiently long baryon lifetime, neither of which can be explained by anthropic arguments, are overall very rare~\cite{Banks:2003es}. In intersecting D-brane models,
it has been estimated that vacua with SM-like gauge group with three generations are one in a billion~\cite{Gmeiner:2005vz}. It would be very interesting to explore whether
such desirable phenomenological features, while rare in the realm of all possible hospitable vacua, may be more common for vacua in funnel-like
regions of the landscape.

The dimensionality of space is another phenomenological issue on which our mechanism can be brought to bear. While it is plausible that life in one or two
dimensions is impossible~\cite{Tegmark:1997jg}, arguments against more than three dimensions are less convincing. (For instance,
the stability of planetary orbits does not forbid mm-size extra dimensions.) It is tempting to speculate that funnel-like regions of the landscape correspond to 
a low effective moduli-space dimensionality, particularly in the vicinity of the lowest-energy vacuum. Such effective reduction in
dimensionality is seen in protein landscapes, where the number of conformation paths decreases dramatically near the native state.
Incidentally, we have offered an explanation for the metastability of our vacuum, but why should the Higgs field be the mediator of this
instability, instead of some other scalar field? A decreased effective dimensionality in our neighborhood of the landscape would
explain why it is incumbent on the Higgs field to be the agent of doom.

Another obvious arena where our mechanism promises to offer new insights is the early universe, particularly inflation. At the phenomenological level, the inflationary
paradigm is highly successful --- a phase of approximate de Sitter expansion yields a nearly scale invariant and Gaussian spectrum of density perturbations,
in excellent agreement with observations. The debate begins the moment one writes down a potential, with the usual points of contention revolving around the 
inflationary potential and/or initial conditions being unnatural/fine-tuned. A tantalizing possibility in the context of our mechanism is that slow-roll inflation
with sufficient number of e-folds, and the near-critical dynamics it entails, occurs generically in funnel-like regions in the approach to the lowest-lying vacuum.
Relatedly, the appearance of the Page time as the optimal lifetime vacua in our analysis is enticing and calls for a deeper understanding. 

Alternatively our mechanism may suggest new ways of realizing inflation. The standard inflationary scenario is based on near-equilibrium dynamics,
as the inflaton equation of motion is derivable from a local Hamiltonian. From this point of view designing a sufficiently flat potential is akin to an
experimentalist dialing the temperature of an Ising spin system near criticality. In contrast, self-organized criticality, and the related notion of generic
scale invariance~\cite{GSI1,GSI2} are intrinsically dissipative, non-equilibrium phenomena. Within our framework it is natural to consider non-equilibrium
inflationary dynamics. An optimal region is an open system, continuously exchanging probability with its environment.
It will be interesting to see whether inflation could be realized as a self-organized critical phenomenon. 

\vspace{.4cm}
\noindent
{\bf Acknowledgements:} We warmly thank Riccardo Penco for collaboration in the early stages of this project and for many discussions. We thank Stephon Alexander, Vijay Balasubramanian, Cliff Burgess, Ulf Danielsson, Frederik Denef, Ben Freivogel, Jaume Garriga, Jonathan Heckman, Thomas Hertog, Mark Hertzberg, Lam Hui, Austin Joyce, Daniel Kabat, Tom Lubensky, Patrick Meade, Yasunori Nomura, Leonardo Senatore, Mark Trodden, Thomas Van Riet and Alex Vilenkin for helpful discussions. We are grateful to Iain Mathieson of the Penn Genetics Department for valuable discussions on natural selection. J.K. is supported in part by the US Department of Energy (HEP) Award DE-SC0013528, NASA ATP grant 80NSSC18K0694, the Charles E. Kaufman Foundation of the Pittsburgh Foundation, and a W.~M.~Keck Foundation Science and Engineering Grant. O.P. is supported by the Simons Foundation (\# 385592, Vijay Balasubramanian) through the It From Qubit Simons Collaboration, and the US Department of Energy contract \# FG02-05ER-41367.

\appendix

\section*{Appendix: Alternate Definition of First Passage Probability}\label{App:FPP}

In the main text we defined the first passage probability $F_{ki}(t)$ as the probability density that one \textit{starts} at node $i$ at $t= 0$ and ends at node $k$ at time $t$. Here we consider a slightly different version, the \textit{conditional first passage probability} $F_{ki}^{(0)}$, defined as the probability density that one \textit{leaves} the node $i$ at $t= 0$ and ends at $k$ within a time interval at time $t$. To see how the two quantities are related, recall~\eqref{PFt}:
\be
P_{ki}(n) = \delta_{ki}\delta_{n,0} + \sum_{m=0}^n F_{ki}(n-m)P_{kk}(m)\Delta t\,. 
\ee
On the other hand, we can alternately write
\be \label{A1}
P_{ki}(n) = \delta_{ki}\delta_{n,0} + \sum_{m=0}^n \sum_{r=0}^{n-m} R_{ii}(r) F^{(0)}_{ki}(n-m-r)P_{kk}(m)\Delta t\,,
\ee
where $R_{ii}(r) = (1- \kappa_i \Delta t)^{r}$ is the probability of sticking to the node $i$ for $r$ time steps. Clearly, 
\be \label{A2}
F_{ki}(n) = \sum_{r=0}^{n} R_{ii}(r) F^{(0)}_{ki}(n-r)\,.
\ee
We can take a Laplace transform of~\eqref{A2} to obtain
\be \label{A3}
\tilde{F}_{ki}(s) =   \frac{1}{\Delta t} \tilde{R}_{ii}(s) \tilde{F}^{(0)}_{ki}(s)\,,
\ee
where
\be
\tilde{R}_{ii}(s) = \sum_{n = 0}^{\infty} R_{ii}(n) e^{-ns\Delta t} \Delta t = \frac{\Delta t}{1 - (1- \kappa_i\Delta t)e^{-s\Delta t}}\,.
\ee
In the $s \to 0$ limit,~\eqref{A3} becomes
\be
\tilde{F}_{ki}(0) =  \frac{1}{\kappa_i \Delta t} \tilde{F}^{(0)}_{ki}(0)\,. 
\ee
The escape probability in~\eqref{escape} can now be written in terms of the conditional first passage probability as
\be
\lim_{t\to \infty} S_{ii}(t) = \frac{\kappa_i\Delta t -  \tilde{F}^{(0)}_{ki}(0)}{{\kappa_i \Delta t} }\,.
\ee
We can interpret the numerator as the probability of leaving the node $i$ at $t=0$ minus the probability of leaving the node $i$ at $t=0$ and returning to $i$ at some later time. This is precisely the probability of leaving the node $i$ at $t=0$ and never returning. Therefore, if we start with some number of walkers at node $i$, the escape probability is the the ratio of the number of walkers who leave at time $t= 0$ and never return to the total number of walkers who leave at time $t=0$.

\end{document}